\documentclass[fleqn,usenatbib]{mnras}

\usepackage{newtxtext,newtxmath}

\usepackage[T1]{fontenc}
\usepackage{soul}
\DeclareRobustCommand{\VAN}[3]{#2}
\let\VANthebibliography\thebibliography
\def\thebibliography{\DeclareRobustCommand{\VAN}[3]{##3}\VANthebibliography}

\usepackage{graphicx}	
\usepackage{amsmath}	
\usepackage[print-unity-mantissa = false]{siunitx} 
\usepackage{pdflscape}
\usepackage{multirow} 
\AtBeginDocument{\RenewCommandCopy\qty\SI}
\usepackage[dvipsnames]{xcolor}

\title[Velocity-dependent SIDM mergers]{Simulations of galaxy cluster mergers with velocity-dependent, rare, and frequent self-interactions}

\author[Sabarish V. M. et al.]{
V. M. Sabarish,$^{1}$\thanks{sabarish.venkataramani@uni-hamburg.de}
Marcus Br\"uggen,$^{1}$
Kai Schmidt-Hoberg,$^{2}$
Moritz S. Fischer$^{3,4}$
and Felix Kahlhoefer$^{5}$
\\
$^{1}$Hamburger Sternwarte, Universität Hamburg, Gojenbergsweg 112, D-21029 Hamburg, Germany\\
$^{2}$Deutsches  Elektronen-Synchrotron  DESY,  Notkestr.~85,  D-22607  Hamburg,  Germany\\
$^{3}$Universit\"ats-Sternwarte, Fakult\"at für Physik, Ludwig-Maximilians-Universit\"at M\"unchen, Scheinerstr. 1, D-81679 M\"unchen, Germany\\
$^{4}$Excellence Cluster ORIGINS, Boltzmannstrasse 2, D-85748 Garching, Germany\\
$^{5}$Institute for Theoretical Particle Physics (TTP), Karlsruhe Institute of Technology (KIT), D-76128 Karlsruhe, Germany\\
}

\date{Accepted XXX. Received YYY; in original form ZZZ}

\pubyear{2023}

\begin{document}
\newcommand{\lcdm}{\Lambda{\rm CDM}}
\newcommand{\solmass}{\mathrm{M_{\odot}}}
\newcommand{\kpc}{\mathrm{kpc}}
\newcommand{\addref}{(e.g. {\color{red} add references})}
\newcommand{\sabc}[1]{{\bf \color{Green}[Sab : #1]}}
\newcommand{\sab}[1]{{\color{Green}#1}}
\newcommand{\review}[1]{#1}
\newcommand{\reviewA}[1]{{#1}}
\newcommand{\msf}[1]{{\color{Emerald} #1}}
\newcommand{\msfc}[1]{{\color{Emerald} \textit{MSF: #1}}}
\newcommand{\som}{\sigma_{0m}}
\newcommand{\cmg}{{\rm cm^2 / g }}
\newcommand{\dd}{\mathrm{d}}
\newcommand{\expval}[1]{\left\langle#1\right\rangle}
\newcommand{\seff}{\sigma_{\rm eff}}
\newcommand{\soneD}{\sigma_{\rm 1D}}
\newcommand{\ksh}[1]{\textcolor{Fuchsia}{{[\bf KSH: #1]}}}
\newcommand{\fk}[1]{\textcolor{magenta}{{[\bf FK: #1]}}}
\renewcommand{\eqref}[1]{equation (\ref{#1})}
\label{firstpage}
\pagerange{\pageref{firstpage}--\pageref{lastpage}}
\maketitle

\begin{abstract}
Self-interacting dark matter (SIDM) has been proposed to solve small-scale problems in $\rm{\Lambda CDM}$ cosmology. In previous work, constraints on the self-interaction cross-section of dark matter have been derived assuming that the self-interaction cross-section is independent of velocity. However, a velocity-dependent cross-section is more natural in most theories of SIDM. \reviewA{Using idealized $N$-body simulations without baryons, we study merging clusters with velocity-dependent SIDM.} In addition to the usual rare scattering in the isotropic limit, we also simulate these systems with anisotropic, small-angle (frequent) scatterings. \reviewA{We find that the collision-less brightest cluster galaxy (BCG) has an offset from the DM peak that grows at later stages.} Finally, we also extend the existing upper bounds on the velocity-independent, isotropic self-interaction cross-section to the parameter space of rare and frequent velocity-dependent self-interactions by studying the central densities of dark matter-only isolated haloes.  For these upper-bound parameters, \review{the DM-BCG offsets just after the first pericentre in the dark matter-only simulations are found to be}  $\lesssim \SI{10}{\kpc}$. On the other hand, because of BCG oscillations, we speculate that the distribution of BCG offsets in a relaxed cluster \review{is a} statistically viable probe. Therefore, this motivates further studies of BCG off-centring in hydrodynamic cosmological simulations.


\end{abstract}



\section{Introduction}

Cold dark matter (CDM) is a fundamental component of the standard ${\rm \Lambda CDM}$ cosmology. It plays a vital role in explaining the formation of the large-scale structure of the universe and the anisotropies in the cosmic microwave background. While cosmological $N$-body simulations within $\lcdm$ have successfully reproduced many observations of the large-scale structure, there seem to be discrepancies between observations and simulations on small scales (see \cite{bullockSmallScaleChallengesLambda2017} for a review of the small-scale problems). A solution to the small-scale problems was proposed by \cite{spergelObservationalEvidenceSelfInteracting2000} via a model of DM, where DM particles can non-gravitationally scatter off each other. Constraints on the self-interaction cross-section can be obtained by studying different astrophysical systems. In particular, relaxed galaxy clusters \citep[e.g.,][]{sagunskiVelocitydependentSelfinteractingDark2021,andradeStringentUpperLimit2021} have provided the most stringent constraint on the cross-section. We also have constraints from galaxy cluster mergers \citep{randallConstraintsSelfInteractionCross2008,harveyNongravitationalInteractionsDark2015}. For a review on astrophysical constraints on self-interacting dark matter (SIDM) see \cite{adhikariAstrophysicalTestsDark2022}. 

Velocity-dependent anisotropic cross-sections are natural in most theories of SIDM \citep[for a review of SIDM models see][]{tulinDarkMatterSelfinteractions2018}. Examples \review{of such models} include light mediator models \citep{tulincollisionlessDarkMatter2013,ackermanDarkMatterDark2009a},  atomic DM \citep{clineScatteringPropertiesDark2014}, strongly interacting DM \citep{boddySelfinteractingDarkMatter2014}. Moreover, velocity-dependent cross-sections are also well motivated observationally. Most constraints on the self-interaction cross-section per unit mass of DM particle ($\sigma/m_\chi)$ in the literature, have been derived assuming velocity-independent and isotropic scattering. For example, \cite{sagunskiVelocitydependentSelfinteractingDark2021} quote an upper limit of $\sigma/m_\chi < \SI{0.35}{\cm\squared\per\g} (95\% \mathrm{C.L.})$ at cluster scales and $\sigma/m_\chi < \SI{1.1}{\cm\squared\per\g} (95\% \mathrm{C.L.})$ at group scales. On the galactic scales, \cite{renReconcilingDiversityUniformity2019} \review{find that $\sigma/m_\chi$ is required to be in the range 3-10$ \ \SI{}{\cm\squared\per\g}$ to explain the observed diversity} in the rotation curves in the SPARC data set. Similarly, \cite{sankarrayConstraintsDarkMatter2022} quote an upper-bound of $\sigma/m_\chi<\SI{9.8}{\cm\squared\per\g} (95 \% \mathrm{C.L.})$. At the scale of dwarf galaxies, there is no concrete upper bound on the cross-section. A cross-section with $\sigma/m_\chi > \SI{30}{\cm\squared\per\g}$ is favoured by the observed central densities of Milky Way's dwarf spheroidal galaxies \citep{correaConstrainingVelocitydependentSelfinteracting2021}. \cite{elbertCoreFormationDwarf2015} find that $\sigma/m_\chi$ can be as large as \SI{50}{\cm\squared\per\g} at these scales and still be consistent with observations. These considerations highlight the viability of velocity-dependent cross-sections. 

Observationally probing the angular dependence is a daunting task. One reason being that the effects of angular dependence are not strong enough when studying the evolution of systems that do not have a preferred direction. For example, \cite{robertsonCosmicParticleColliders2017} and \cite{fischerNbodySimulationsDark2021a} simulated isolated haloes using $N$-body simulations and found that there is no big difference in the evolution of core-sizes between isotropic and anisotropic cross-sections for a given choice of parameters. Moreover, simulating differential cross-sections which peak for tiny scattering angles, using conventional SIDM implementations \citep[e.g.][]{rochaCosmologicalSimulationsSelfinteracting2013} is prohibitively expensive. This type of interaction will be called frequent self-interactions \citep[as introduced in][]{kahlhoeferCollidingClustersDark2014} as opposed to rare self-interactions corresponding to large-angle scattering.  Frequently self-interacting dark matter (fSIDM) is natural in the mass-less mediator limit of light mediator models. The scattering of DM particles in this regime can be modelled as a drag force, where the drag force depends not on the total cross-section but on the momentum transfer cross-section given by~\cite{kahlhoeferDarkMatterSelfinteractions2017} and \cite{robertsonCosmicParticleColliders2017}
\begin{equation}\label{eqn:def:momentum transfer cross-section}
\sigma_\mathrm{T}=2 \pi \int_{-1}^{1} \frac{\mathrm{d} \sigma}{\mathrm{d} \Omega_{\mathrm{cms}}}\left(1-|\cos \theta_{\mathrm{cms}}| \right) \mathrm{d} \cos \theta_{\mathrm{cms}} .
\end{equation}
\review{where $\theta_{\rm cms}$ and $\Omega_{\rm cms}$ are the scattering angle, and solid angle in the centre of mass frame}. Frequent self-interactions have previously been studied by \cite{kahlhoeferCollidingClustersDark2014},\cite{kummerEffectiveDescriptionDark2018,kummerSimulationsCoreFormation2019},\cite{fischerNbodySimulationsDark2021a,fischerUnequalmassMergersDark2022} assuming velocity independence. 

Mergers of galaxy clusters are interesting test beds for models of SIDM since the mass distribution of the system could be measured through lensing. The gas and galaxies can be probed through their direct emission in various wavelengths. The existence of offsets between the DM component and galaxies may hint at DM self-interactions \citep{randallConstraintsSelfInteractionCross2008}. Moreover, mergers are sensitive to both velocity and angular dependence of the scattering cross-section. First, as the haloes undergo many pericentre passages, the collisional velocity changes with time. Scattering velocities are the largest at the first pericentre passage and, subsequently, the haloes slow down with every passage. The self-interactions at the pericentre passages are mainly responsible for an increase in the offset. Thus, the evolution is sensitive to the parameters of velocity-dependent cross-section. Secondly, mergers unlike isolated haloes also have a preferred direction, i.e.\ the merger axis. \cite{fischerUnequalmassMergersDark2022} find that offsets are larger for frequent self-interactions with a given $\sigma_\mathrm{T}$, when compared to rare self-interactions of the same $\sigma_\mathrm{T}$. They also showed that small-angle scattering can produce larger offsets than the maximal possible offset from isotropic scattering.


There have been earlier studies that have simulated mergers. For example, studies with velocity-independent isotropic cross-sections have been done by \cite{kimWakeDarkGiants2017} who simulated equal-mass mergers ; \cite{robertsonWhatDoesBullet2017} simulated a bullet cluster like system. \cite{fischerNbodySimulationsDark2021a,fischerUnequalmassMergersDark2022} studied equal and unequal-mass mergers with isotropic and anisotropic velocity-independent cross-sections. \cite{robertsonCosmicParticleColliders2017} looked at mergers until just after the first pericentre passage. They used a velocity-dependent isotropic cross-section and a velocity-dependent cross-section that corresponds to Yukawa scattering under the Born approximation. It is unknown as to how the merger evolution is affected at late stages by velocity-dependent self-interactions. Similarly, mergers in the fSIDM regime with velocity dependence are yet to be studied.
 
In this work, we aim to  \textit{(i)} study the qualitative differences in merger simulations between velocity-dependent and independent cross-sections, \textit{(ii)} extend the upper bound on constant cross-section quoted by \cite{sagunskiVelocitydependentSelfinteractingDark2021} to the parameter space of velocity-dependent cross-section, \textit{(iii)} find the maximum offsets between DM and the brightest cluster galaxy (BCG) with velocity-dependent cross-section parameters that are consistent with upper bound parameters. To this end, we simulate
the full evolution of galaxy cluster mergers and isolated haloes with rare and frequent self-interactions. In a companion paper \citep{fischer2023d}, cosmological simulations are studied with velocity-dependent fSIDM. The paper is presented as follows. In section~\ref{sec:methods} we briefly describe our numerical scheme and the SIDM models that are considered. In section~\ref{sec:varying w} we present our simulation results, which illustrate the qualitative differences between velocity-dependent and velocity-independent cross-section simulations. In section~\ref{sec:results_cdens_matched}, we describe the simulations of mergers with cross-section parameters that correspond to the 95\% C.L. limits provided in \cite{sagunskiVelocitydependentSelfinteractingDark2021}. In section~\ref{sec:conclusion}, we summarize our results and conclude. Additional details are provided in the appendices \ref{app:validating resolution} and \ref{app:central density rescaling}.

\section{Methods} \label{sec:methods}

In this section, we describe the numerical setup of our simulations, and we discuss our choice for the self-interaction cross-section that is used in the simulations.

\subsection{Numerical method} \label{sec:numerical_method}
For our simulations, we use the cosmological $N$-body simulation code \review{\textsc{OpenGadget3}}, adapted for frequent self-interactions using the implementation given in \cite{fischerNbodySimulationsDark2021a}. In this section, we briefly describe the implementation of rare and frequent self-interactions. The numerical scheme for the self-interactions of rarely self-interacting dark matter (rSIDM) follows \cite{fischerNbodySimulationsDark2021a}, which is similar to the method described in \cite{rochaCosmologicalSimulationsSelfinteracting2013}. \review{In this scheme, the probability that a numerical particle 1 with mass $m_1$ scatters off another numerical particle 2 with mass $m_2$ is given by,  
\begin{equation}\label{eqn : scattering probability}
    P_{12} = \frac{\sigma(v_{12})}{m_\chi} \ m_2 \ |v_{12}| \ \Delta t  \ \Lambda_{12} ,
\end{equation}}
where $v_{12}$ is the relative velocity between the numerical particles 1 and 2, $\Delta t$ is the time-step used in the simulation, $\Lambda_{12}$ is the kernel overlap integral, $\sigma\review{(v_{12})}/m_\chi$ is the total cross-section per unit mass of DM particle. For more details on the implementation and the choice for the kernel, see appendix A and B of \cite{fischerNbodySimulationsDark2021a}. A collection of kernels used in other modern implementations of SIDM within $N$-body simulations can be found in \citet[equations 11-15]{adhikariAstrophysicalTestsDark2022}.

The drag force of frequent self-interactions is based on the relation derived in \cite{kahlhoeferCollidingClustersDark2014}, which describes the deceleration rate experienced by a particle as it travels through a constant background density of DM,
\begin{equation}
    R_{\rm dec} = \frac{1}{v_0} \frac{\dd v_{\|}}{\dd t } = \frac{\rho_0 v_0 \sigma_\mathrm{T}}{2 m_\chi},
\end{equation}
$v_0$ is the velocity of the particle, \review{$v_\|$ is the parallel component of the velocity of the particle,} $\rho_0$ is the background density, $\sigma_\mathrm{T}$ is the momentum transfer cross-section defined in \eqref{eqn:def:momentum transfer cross-section}. We can also see that the deceleration rate captures the rate of change of the parallel component of the velocity. Therefore, the above expression can then be cast into an expression for drag force for the physical particles. \review{The drag force as experienced by numerical particles, labelled 1 and 2, in the $N$-body code is given as} \citep{fischerNbodySimulationsDark2021a},
\review{
\begin{equation}\label{eqn:drag force}
    F_{\rm drag} = \frac{1}{2} | v_{12}|^2 \frac{\sigma_\mathrm{T}(v_{12})}{m_\chi} m_1 m_2 \Lambda_{12},
\end{equation}
}
which is proportional to the momentum transfer cross-section $\sigma_\mathrm{T}$. \review{From here on, we will drop the subscripts and let $v$ denote the relative velocity between DM particles.}

\review{In addition to the drag force, momentum is added to the particles in a random but perpendicular direction relative to the motion in the centre-of-mass frame. As for rSIDM, the post-scattered momenta in the centre-of-mass frame have the same absolute value as the pre-scattered ones.}

\review{In our SIDM implementation, we search for pairs of potentially interacting particles.
A search for the neighbours of each particle achieves this, employing the same tree structure as used in the gravity computation.

The code employs an adaptive time-stepping scheme, where the time-step of an individual particle is obtained by the minimum of different time-step criteria, i.e.\ given by gravity and self-interactions.

The implementation of rSIDM and fSIDM does conserve momentum as well as energy explicitly.
This is achieved by executing the scattering computations for pairs of particles in a consecutive manner such that their velocities are updated after each scattering event and used for the next one. At the same time, we allow for multiple interactions per particle per time-step. 

Massively parallel computations of the self-interactions are enabled by a parallelized implementation using the message passing interface (MPI). Note, here we order the communication and computation of the processes in a manner to reduce waiting time.}

The implementation of the velocity-dependent self-interactions into the SIDM module has been described in detail by \cite{fischer2023d}.  To ensure numerically stable results, a novel time-step criterion has been added. This criterion is based on the velocity at which self-interactions are strongest, i.e.\ on the maximum of $\sigma_\mathrm{T}(v) \, v$, instead of the velocity distribution that an individual particle encountered in the previous time-step. In principle, such a scheme guarantees that the simulation time-step is always sufficiently small to account for $\sigma_\mathrm{T}(v)$ for any $v$.

\subsection{Initial conditions and simulation parameters}

In this work, we simulate both DM-only isolated haloes and galaxy cluster merger with two different merger mass ratios (MMR). Firstly for the isolated DM-only haloes, they have a virial mass of $\SI{1e15}{\solmass}$ and are initialized with an NFW density profile \cite{navarroStructureColdDark1996a}, i.e.\
\begin{equation}
    \rho(r) = \frac{\delta_c \rho_{\rm{crit}}}{(r / r_s) (1 + r / r_s)^2} := \frac{\rho_0}{(r / r_s) (1 + r / r_s)^2}.
\end{equation}
DM positions are sampled from a probability density function such that the density profile follows the NFW profile. Given the virial mass of the halo and the critical density $\rho_{\rm crit}$ of the universe, the parameters of the NFW profile are determined as follows: \textit{(i)} the concentration parameter $c_{\rm vir}$ is determined using the concentration-mass relation given in \cite{duttonColdDarkMatter2014a}, \textit{(ii)} characteristic density $\delta_c$ is computed using $c_{\rm vir}$ from the earlier step \citep[using Eq.2 of][]{navarroStructureColdDark1996a}, \textit{(iii)} scale radius $r_s$ is computed using its definition $r_s = r_{\rm vir}/c_{\rm vir} $. The NFW parameters thus obtained are given in \autoref{tab:merger nfw params}.  Using the resulting density profile, the initial velocity dispersion $\expval{v^2}_{\rm ini}(r)$ is obtained by integrating the Jeans equation \citep{binneyGalacticDynamicsSecond2008}. Then, initial velocities in each radial bin are drawn from a Gaussian distribution with the variance $\expval{v^2}_{\rm ini}(r)$. The isolated DM halo simulations use the same NFW parameters as the main halo of the merger for initial conditions (ICs). We simulate mergers with MMR $\in \lbrace 1,5 \rbrace$. The barycentre of the clusters are initially \SI{4000}{kpc} apart and they are put on course towards each other with a relative velocity of \SI{1000}{\km\per\s} as summarized in \autoref{tab:merger ics}.

In the galaxy cluster used in the merger simulations, the cluster has three particle species: DM, galaxy and BCG. Galaxies and BCG are approximated to be collision-less, while DM is collisional with self-interaction characterized by its cross-section. An equal number of DM and galaxy particles is used in the simulation. A sufficient number of galaxy particles are chosen to ensure that it is easier to find the peak position of the galaxies. The main halo has a virial mass of $10^{15} \solmass$ and both species initially follow an NFW profile. The particle masses are as follows: for DM, $m_{\rm DM} = 2\times10^9 \ \solmass$, for galaxies, $m_{\rm Gal} = 4\times10^7 \ \solmass$. In addition, the brightest cluster galaxy (BCG) is represented by a single particle at the centre of the halo with a mass $m_{\rm BCG} = 7\times10^{11} \ \solmass$. As the BCG is approximated to be a point particle, the effects of gravitational scattering become strong if the mass is large, therefore the BCG particle is taken to be less massive than the observed BCGs. This choice is adopted from earlier studies \citep[e.g.][]{fischerUnequalmassMergersDark2022,kimWakeDarkGiants2017}. \reviewA{The mass resolution chosen works reasonably well for measuring offsets in simulation since the simulation is run only for a few dynamical time-scales. The detailed effects of such a treatment on measurements other than offsets are yet to be studied.}
We use a fixed gravitational softening length of $\epsilon$ = \SI{1.2}{kpc} for all 
particles. For both mergers and isolated haloes we use an adaptive kernel size for the DM self-interactions, such that the number of neighbours within each particles’ kernel, $N_{\rm ngb}$ is equal to 64. This choice follows from \cite{fischerNbodySimulationsDark2021a}.

All simulations have been performed with a resolution of $\mathcal{O}(10^6)$ particles. For certain cross-section parameters, the simulations were rerun at a higher resolution of $\mathcal{O}(10^7)$ particles to validate the lower resolution runs (see Appendix \ref{app:validating resolution}).

\begin{table}
	\centering
	\caption{This table contains the NFW parameters used in generating the ICs for merger simulations. The first column contains the virial mass, the second the density parameter $\rho_{0}:=\delta_{c} \ \rho_{\rm crit}$, followed by scale radius $r_s$, \review{concentration parameter} and  number of DM and galaxy particles. DM-only isolated haloes have the same NFW parameters as given in the first row. 
 }
	\label{tab:merger nfw params}
    \renewcommand{\arraystretch}{1.5} 
    \begin{tabular}{c|c|c|c|c|c}
      M$_{\text{vir}}$ & $\rho_0 $ & $r_s$ & $\review{c}$ & $N_{\rm DM} = N_{\rm Gal}$ & $N_{\rm BCG}$\\
      $(\solmass)$&$(\SI{}{\solmass \kpc^{-3}})$& $(\SI{}{\kpc})$&\\
      \hline
      $10^{15}$ & $1.33\times10^6 $ & $389.3$ & \review{5.42} & 1009878 & 1 \\
      $\SI[exponent-product = \times]{2e14}{}$ & $\num[exponent-product = \times]{1.908e6}$  & 194.76 & \review{6.33} &  181319 & 1
    \end{tabular}
\end{table}

\begin{table}
	\centering
	\caption{\reviewA{This table contains the initial separation,  initial relative velocity between the two clusters and the relative velocity of the BCGs at the first pericentre passage.}}
	\label{tab:merger ics}
    \renewcommand{\arraystretch}{1.5} 
    \begin{tabular}{c|c|c|c}
      Merger mass ratios & $x_{\rm ini}$ & $\Delta v_{\rm ini}$  &\reviewA{$\Delta v^{(\rm BCG)}_{\rm FPP}$}\\
      $(\solmass : \solmass)$  & $(\kpc)$ & $(\unit{\km\per\s})$  & \reviewA{$(\unit{\km\per\s})$} \\
      \hline
      $10^{15} : 10^{15} $ & $ 4000 $ & $ 1000 $  & \reviewA{$5500$}\\
      $10^{15} : 2\times10^{14} $ & $ 4000 $ & $ 1000 $ & \reviewA{$4800$} 
    \end{tabular}
\end{table}

\subsection{Dark matter cross-section model}\label{sec:dm model}

We assume a fairly generic form for the self-interaction momentum transfer cross-section for rare and frequent self-interactions, parametrized as follows \citep{gilmanStrongLensingSignatures2021,yangGravothermalSolutionsSIDM2023},
\begin{equation}\label{eqn:momentum transfer cross-section used in sims}
    \dfrac{\sigma_\mathrm{T}\review{(v)}}{m_\chi} = \frac{\sigma_0}{m_\chi} \left({1+\frac{v^2}{w^2}}\right)^{-2},
\end{equation}
\begin{figure}
    \centering
    \includegraphics[width=0.45\textwidth]{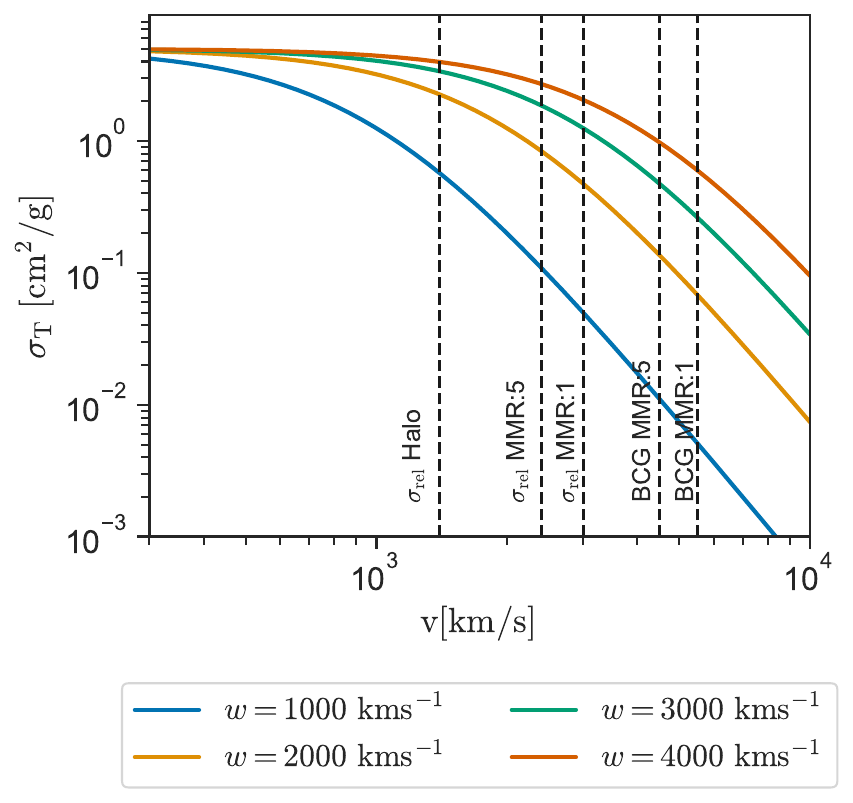}
    \caption{Momentum transfer cross-section for $\som = \SI{5}{\cm\squared\per\g} $\review{, and} the values of $w$ are given in the legend. The vertical coloured dashed lines correspond to different velocity scales in the system. The left most is the \review{relative velocity dispersion} within \SI{100}{kpc} of the $\SI{1e15}{\solmass}$ cluster. The second and third from the left correspond to the \review{relative velocity dispersion} within \SI{100}{kpc} around the barycentre at the first pericentre for MMR:5 and MMR:1 respectively. Finally, the second last and the last are the relative velocity of the BCGs at the first pericentre passage of the system with MMR:5 and MMR:1, respectively.}
    \label{fig:cross-section}
\end{figure}
where $v$ is the relative velocity of the DM particles. 
We furthermore assume that the total cross-section has the same velocity dependence as the momentum transfer cross-section, such that
\begin{equation}
    \frac{\sigma_\mathrm{T}(v)}{\sigma_\mathrm{T}(v=0)} = \frac{\sigma(v)}{\sigma(v=0)} \; . \label{eq:proportionality}
\end{equation}
This assumption is automatically satisfied if the differential cross-section is a separable function, i.e.\ it is of the following form,
\begin{equation}\label{eqn:differential cross-section + generic form}
    \frac{\dd \sigma}{\dd \cos\theta} = N \frac{\sigma_0}{m_\chi}\Theta(\theta) g(w,v),
\end{equation}
where $N$ is a normalization constant, $\Theta(\theta)$ captures the angular dependence and $g(w,v) = (1+v^2/w^2)^{-2}$. 
However, even for non-separable differential cross-sections, such as Rutherford scattering, the assumption in \eqref{eq:proportionality} gives a useful approximation.

Depending on the choice of $N$ and $\Theta(\theta)$, the above \review{differential} cross-section can correspond to either frequent or rare self-interactions. In this work, we consider isotropic scattering for the case of rare self-interactions, such that $\Theta(\theta)$ is simply a number. The total cross-section to calculate the scattering probability in \eqref{eqn : scattering probability} is then given by $\sigma\review{(v)} = 2 \sigma_\mathrm{T}\review{(v)}$, see \eqref{eqn:def:momentum transfer cross-section}.
For frequent self-interactions, on the other hand, $\Theta(\theta)$ is strongly peaked for small angles. The normalization, $N$, is then chosen such that the momentum transfer cross-section given in \eqref{eqn:momentum transfer cross-section used in sims} is recovered, which is used in the simulations to compute the drag force experienced by the particles, see \eqref{eqn:drag force}. 

To run the simulations, the parameters $\som$ and $w$ must be chosen, where $\som := \sigma_0 / m_\chi$. In this paper, we are interested in studying the qualitative differences in the evolution of galaxy cluster mergers between the different regions of $\som-w$ parameter space. A priori, it is conceivable that there are degeneracies in this parameter space, such that different parameter combinations lead to very similar merger observables.  Such a degeneracy was found by  \cite{yangGravothermalSolutionsSIDM2023} (see their fig. 8) when analysing the rotation curve of LSB dwarf galaxy UGC 128. In other words, it was possible to compensate a change in $w$ by an appropriate change in $\som$. We refer to a prescription to determine the cross-section normalization $\som$ for a given velocity dependence (determined by $w$) as \emph{matching}. In the following, we will explore different matching procedures.

The choice for $w$ follows from the typical relative velocity scales. These scales can be estimated by running CDM simulations. The observed values are displayed as dashed lines in \autoref{fig:cross-section}. The largest observed scales are the relative velocities of the infalling BCGs shown as the last two lines corresponding to MMR:5 and MMR:1 respectively. The average scattering velocity with which DM particles scatter off each other within \SI{100}{kpc} around the barycentre at the first pericentre passage \review{are shown}. Finally, the \review{relative velocity dispersion within \SI{100}{\kpc} of a \SI{1e15}{\solmass} halo is displayed as the leftmost vertical dashed line} and it is approximately \SI{1400}{\km\per\s}. This implies that for any value of $w$ larger than \SI{1400}{\km\per\s}, the self-interactions within the halo will be in the weakly velocity-dependent regime. Therefore, the following choice for $w$ is made, $w \in \{ 1000 ,2000,3000,4000 \} \SI{}{\km\per\s}$.

We have analysed cluster mergers with $\som$ chosen according to two matching procedures: The first is to choose the same value of $\som$ for all chosen values of $w$, the second is to choose $\som$ such that the evolution of the central density of the isolated haloes for different values of $w$ is similar. These procedures will be explained further in the following sections.

\subsection{Analysis methods}

In order to find the peak position of any component, i.e.\ DM or galaxies, we use the peak finding method based on the gravitational potential, see \cite{fischerUnequalmassMergersDark2022}. In the simulations, all particles have a unique particle ID assigned to them. Using the ID, particles belonging to a given halo can be identified. Then, the gravitational potential in each cell in a grid is computed. The cell with the lowest potential corresponds to the position of the peak. \cite{fischerUnequalmassMergersDark2022} also propose the isodensity-based peak finding algorithm. In this algorithm, the peaks are identified as the cell in the merger plane with the highest projected density. This method is closer to observations where, for example, gravitational shear measurements can be used to infer mass densities. We find that gravitational potential based peak finding is more reliable when the simulation is run with low resolution. Hence, this is our choice for finding peaks. To find the errors on the peak position, we bootstrap the particle distribution 20 times and determine 20 such peaks. Then we estimate the error, by finding the standard deviation in the obtained peak positions.

We define offsets as the distance between the peaks of two different species of the same cluster. For example, $d_{\rm DM-BCG}$ is the distance between the DM and BCG peak of a given cluster.

\section{Varying \texorpdfstring{$\lowercase{{w}}$}{\textit{w}} only }\label{sec:varying w}

In this scheme, the same value of $\som$ is used for different values of the parameter $w$. Even though it is a very simple choice, it is easier to observe the qualitative difference introduced by velocity-dependent cross-sections. To make the differences stand out, we use a large value of $\som=\SI{5.0}{\cm\squared\per\g}$. 
We confirmed the results to be qualitatively similar but less pronounced for smaller values of $\som$.
The names for the individual runs shown in the plots are tabulated in \autoref{tab:s5p0 labels}.

\begin{table}
\centering
\renewcommand{\arraystretch}{1.5}
\begin{tabular}{c|c|c}
Simulation name & $\som$                     & $w$                \\
                & [\SI{}{\cm\squared\per\g}] & [\SI{}{\km\per\s}] \\
\hline
FC5p0           & 5                          & $\infty$           \\
Fw1000s5p0      & 5                          & 1000               \\
Fw2000s5p0      & 5                          & 2000               \\
Fw3000s5p0      & 5                          & 3000               \\
Fw4000s5p0      & 5                          & 4000        
\end{tabular}
\caption{Simulation labels and the corresponding cross-section parameters. The generic form of labels is XwDsNpM, where X is \review{F(frequent), R(rare), or C(constant)}. D that follows w is the value of the parameter $w$ to be read as D \SI{}{\km\per\s} and NpM following the s is the value of $\som$ to be read as N.M \SI{}{\cm\squared\per\g}. }\label{tab:s5p0 labels}
\end{table}

\begin{figure}
    \centering
    \includegraphics[width=0.48\textwidth]{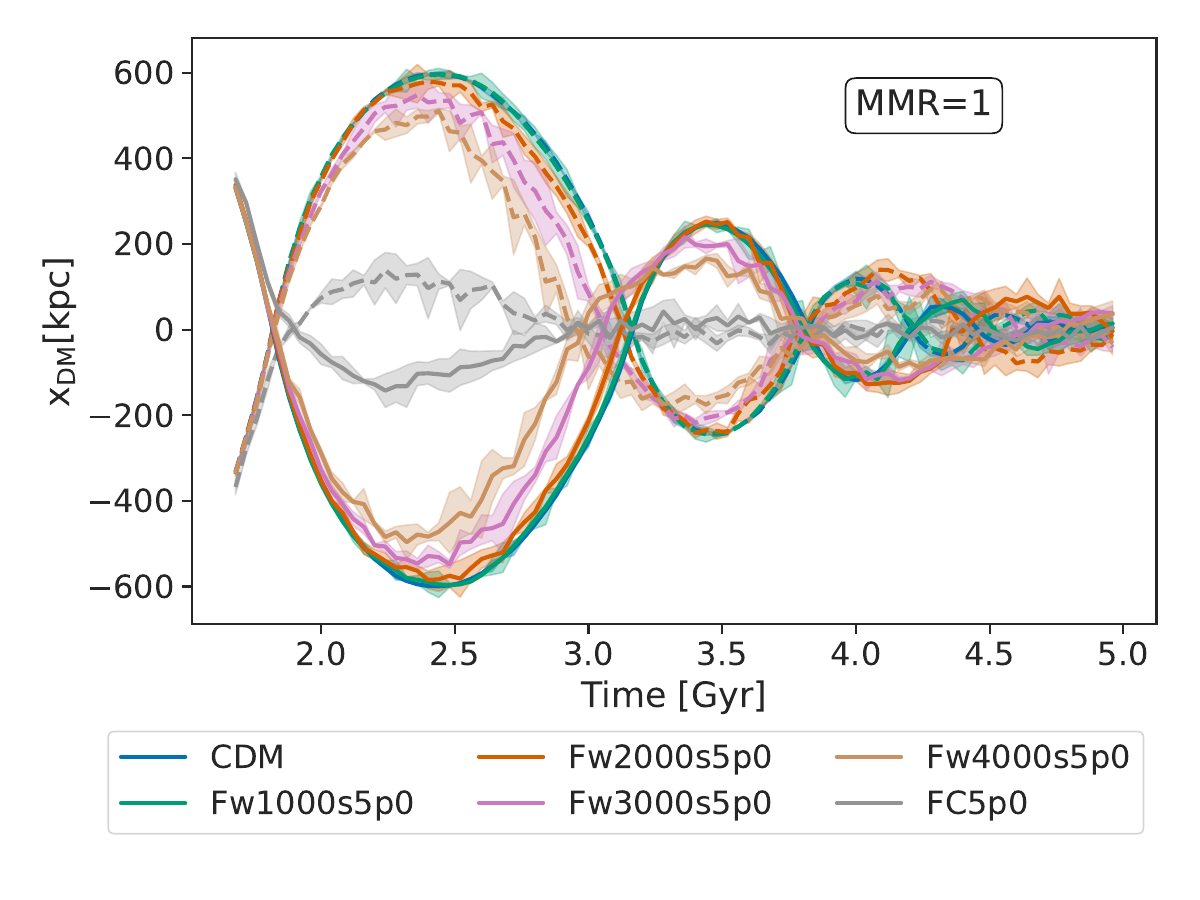}
    \includegraphics[width=0.48\textwidth]{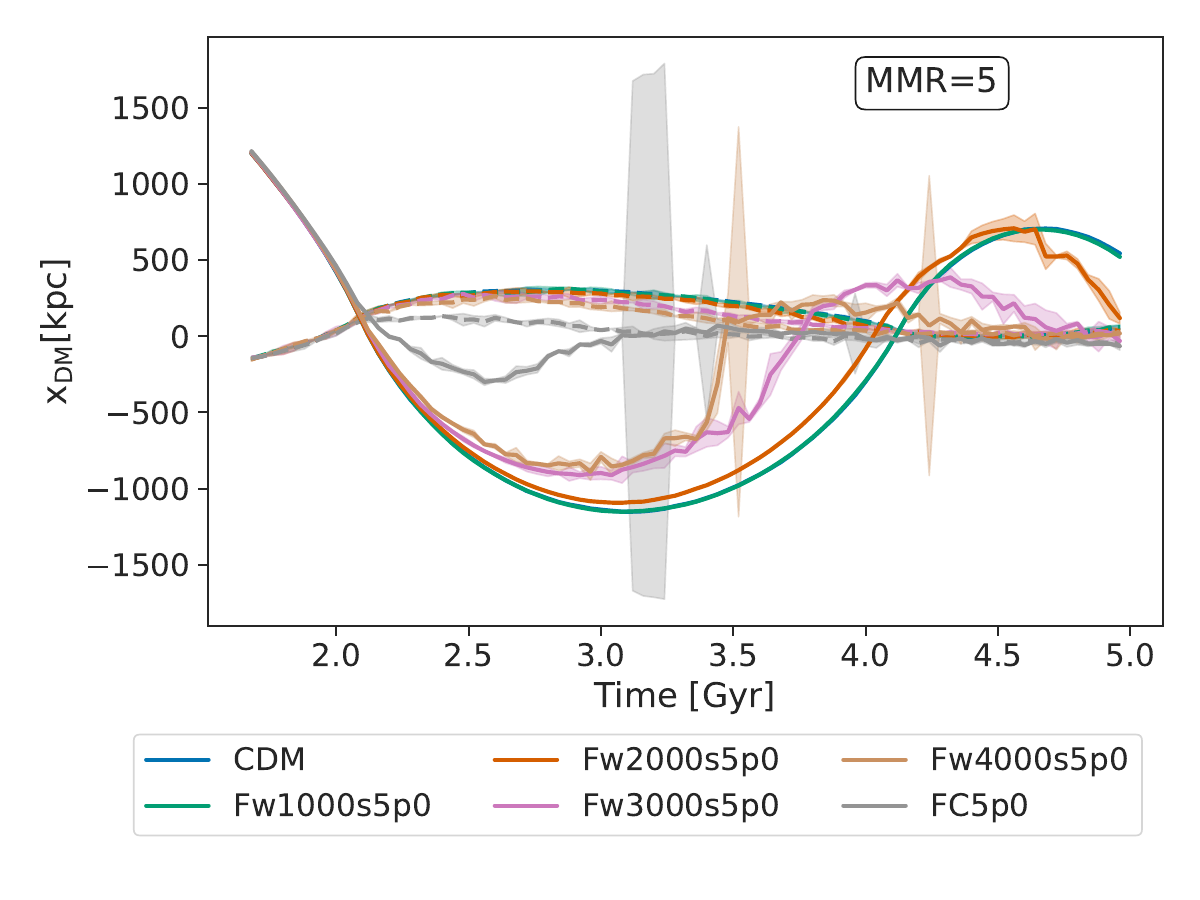}
    \caption{The dotted and solid lines correspond to the DM peak position of the main halo and subhalo respectively. Upper panel: DM peaks in equal-mass merger, lower panel: Merger with mass ratio 5. Plot labels are described in \autoref{tab:s5p0 labels}. Simulation results presented in this figure corresponds to varying values of $w$ and a fixed $\som$. }
    \label{fig:DM peak position 5p0}.
\end{figure}

\subsection{DM peak position}\label{subsec:5p0: DM peaks}

The plots of the peak positions of DM against time for equal and unequal-mass mergers are given in \autoref{fig:DM peak position 5p0}. The peaks of the main halo are marked by dotted lines, while those of subhalo are indicated by solid lines. First we observe that for \review{constant $\sigma_T$} the drag force from self-interaction is strong enough to stop the DM component in their tracks and they coalesce at the first pericentre passage. For the same $\som$, the DM peak positions in the velocity-dependent SIDM runs are closer to the CDM run for smaller values of $w$. This observation matches our expectation that, for fixed $\som$, increasing $w$ increases the effective self-interaction strength. 

The separation of the peaks at the first apocentre is largest for CDM and decreases with increasing self-interaction strength. This can be attributed to the fact that DM particles experience a drag force and thus stay closer to the centre-of-mass of the system.

The first pericentre passage occurs at different times for different parameter combinations. Albeit a small effect, self-interactions tend to increase the time taken to reach the first pericentre passage. Then the second passage occurs earlier for increasing self-interaction strength. For example, we see that the first pericentre passage occurs slightly later \review{ in the simulation with constant $\sigma_T$}. This could be understood as self-interactions generating pressure that resists the infall, thus slowing the halo down. 
The difference is large for \review{constant $\sigma_T$ } because the effective self-interaction strength of the velocity-dependent ones is much smaller than the \review{one of constant $\sigma_T$ } at early stages, as $\sigma_\mathrm{T}(v)$ is small for $v>w$.

At later stages of the evolution, the scenario changes. For example, at the third apocentre, we see that the oscillation in the DM peak for $w=\SI{2000}{\km\per\s}$ has an amplitude that is greater than that of CDM. \review{During these stages, the central density of the haloes are lower for SIDM simulations (see \autoref{fig:M5R1 central density s5p0}). Therefore, the oscillations are less dampened by dynamical friction for SIDM simulations.} 

For the unequal-mass mergers, the subhalo dissolves faster, making it difficult to identify the DM peaks during later stages of the evolution. Both equal and unequal mass mergers have identical initial separation and initial relative velocity. 
Therefore, in unequal mass merger, the less massive cluster traverses more distance than the massive one and this leads to fewer oscillations. For instance, in \autoref{fig:DM peak position 5p0}, we see that within 5 billion years, the subhalo in MMR:5 system has undergone fewer pericentre passages than the equal-mass merger.

\subsection{BCG peak position}\label{sec: bcg oscillations}

\begin{figure}
    \centering
    \includegraphics[width=0.48\textwidth]{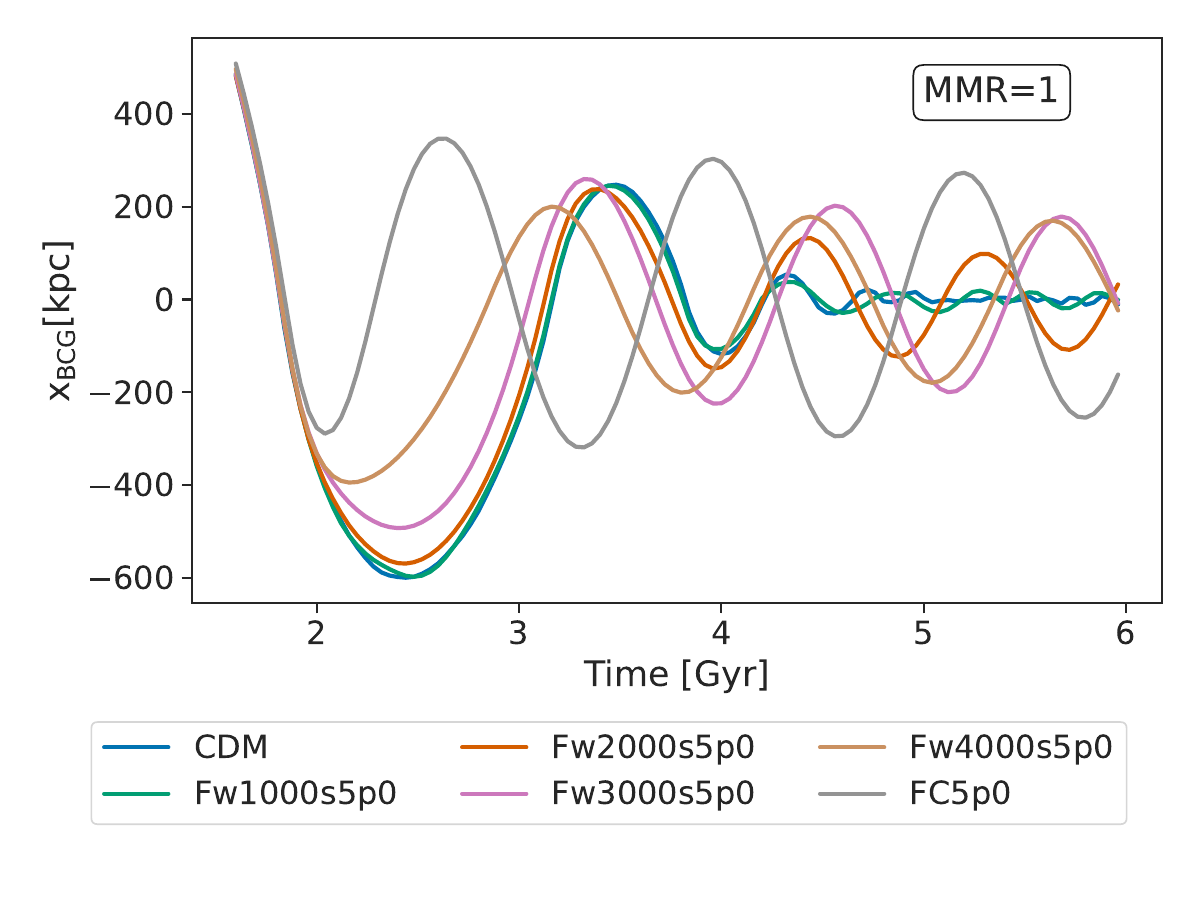}
    \includegraphics[width=0.48\textwidth]{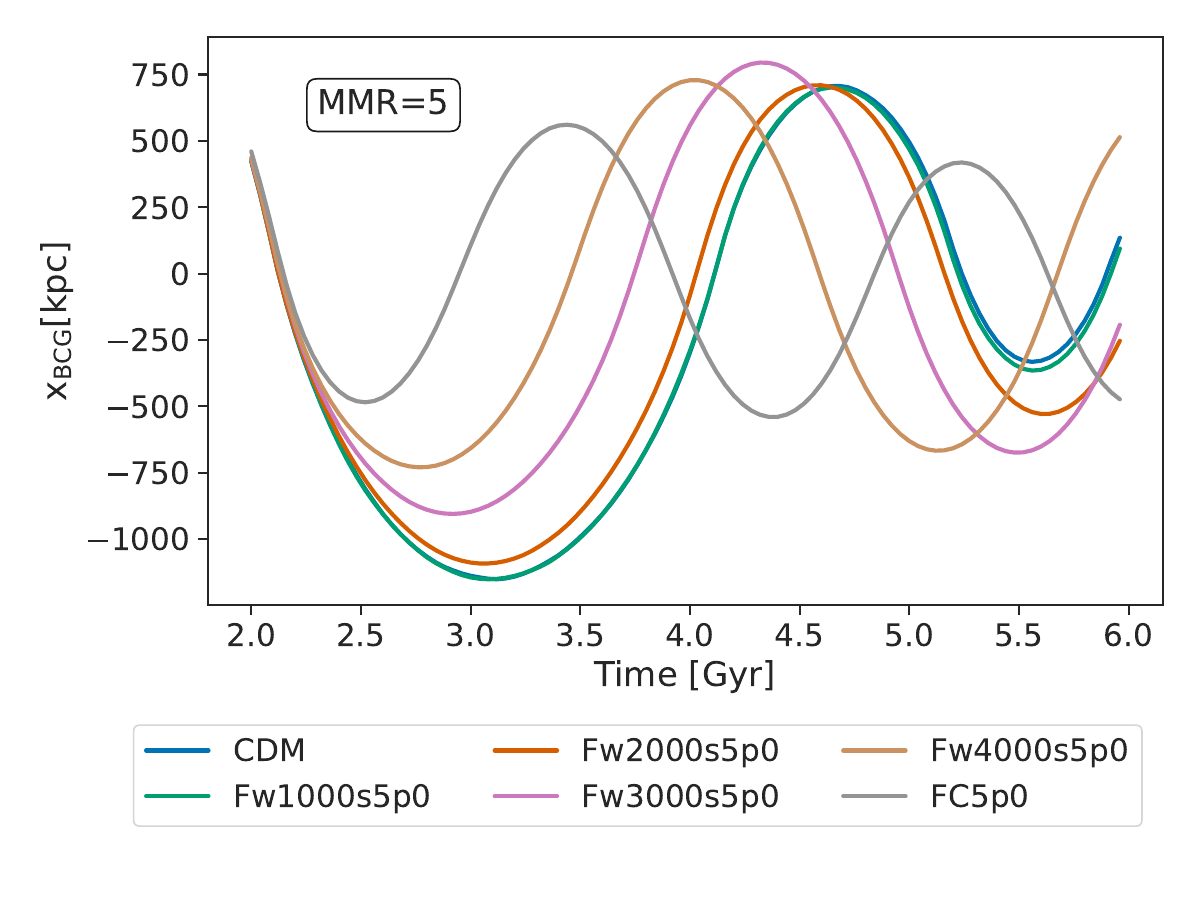}
    \caption{BCG position of the subhalo vs. time. The upper panel shows peak positions in the equal-mass merger, while the lower panel corresponds to the unequal-mass merger. For better visibility, we plot only the position of the BCG of the subhalo. Plot labels are described in \autoref{tab:s5p0 labels}.
}
    \label{fig:BCG peaks 5p0}
\end{figure}

The BCG positions for subhaloes are given in \autoref{fig:BCG peaks 5p0}, the upper panel corresponds to the equal-mass merger system, while the lower panel corresponds to the unequal-mass merger. \review{In CDM simulations, the BCG oscillations are damped faster compared to SIDM simulations.} This general feature has already been observed in earlier work \citep{fischerUnequalmassMergersDark2022,fischerNbodySimulationsDark2021a,kimWakeDarkGiants2017}. This could be explained by noting that the merger remnants have a cored density distribution at the centre owing to the self-interactions. On the other hand, merger remnants in CDM simulations have larger central densities. As a result, the oscillations dampen out faster in CDM due to dynamical friction.

In the equal-mass merger, we observe that the peak positions of subhalo BCGs are closer to the CDM for smaller values of $w$ at the early stages of the merger evolution. At later stages, around 5 billion years, the BCG oscillations in the CDM simulation have dampened considerably. On the other hand, the oscillations approximately stay constant for the constant $\sigma_{\rm T}$ simulation for the period $1-8$ billion years, as shown in \autoref{fig:BCG peaks 5p0}. The position of the BCG in velocity-dependent simulations start deviating from CDM with time. For example, let us consider the $w=\SI{2000}{\km\per\s}$ simulation : \textit{(i)} we see that the curve is initially close to the CDM simulation \textit{(ii)} during the period $4-7$ billion years, the oscillations have approximately a constant amplitude, and they have significantly deviated from the CDM simulation. \review{The typical velocities around the DM peak at various stages of the evolution are given in \autoref{fig:M5R1 veldispr s5p0}.} \review{The central relative velocity dispersion around the DM peak is large at the first pericentre owing to the active merging. At later stages, when the haloes have slowed down, they have a very slowly rising relative velocity dispersion for the rest of the simulated time period. }  Therefore, from \autoref{fig:cross-section} we see that for $w=\SI{2000}{\km\per\s}$ and $v>\SI{2000}{\km\per\s}$, the $\sigma_{\rm T}$ is less than $\SI{1}{\cm\squared\per\gram}$, whereas for $v<\SI{2000}{\km\per\s}$ the cross-section is larger than $\SI{1}{\cm\squared\per\gram}$. Thus, merger remnants experience larger self-interactions at later stages, leading to more cored distributions and lesser dynamical friction. Cumulatively, this leads to steady oscillations at later stages.

The distance travelled by the BCGs at the first apocentre is observed to become smaller with increasing values of $w$. This can be understood by looking at the DM peaks. For example, the DM haloes coalesce at the first pericentre passage for the \review{constant $\sigma_T$ } simulation. As a result, the BCG experiences a larger gravitational force due to the coalesced DM distribution at the barycentre. This accumulation of DM at the barycentre reduces with decreasing $w$ since the average interaction strength reduces with $w$. This effectively leads to smaller amplitudes at the first apocentre in both equal and unequal-mass mergers. Immediately after the second pericentre passage, the amplitude of the velocity-dependent SIDM and CDM curves have decreased significantly due to dynamic friction.  \review{In the constant $\sigma_T$ simulation }, the DM peaks have come to rest, and the merger remnants start coring. This leads to the persistence of the BCG oscillation amplitude.

\subsection{Morphology}

\begin{figure*}
    \centering
    \includegraphics[width=0.8\textwidth]{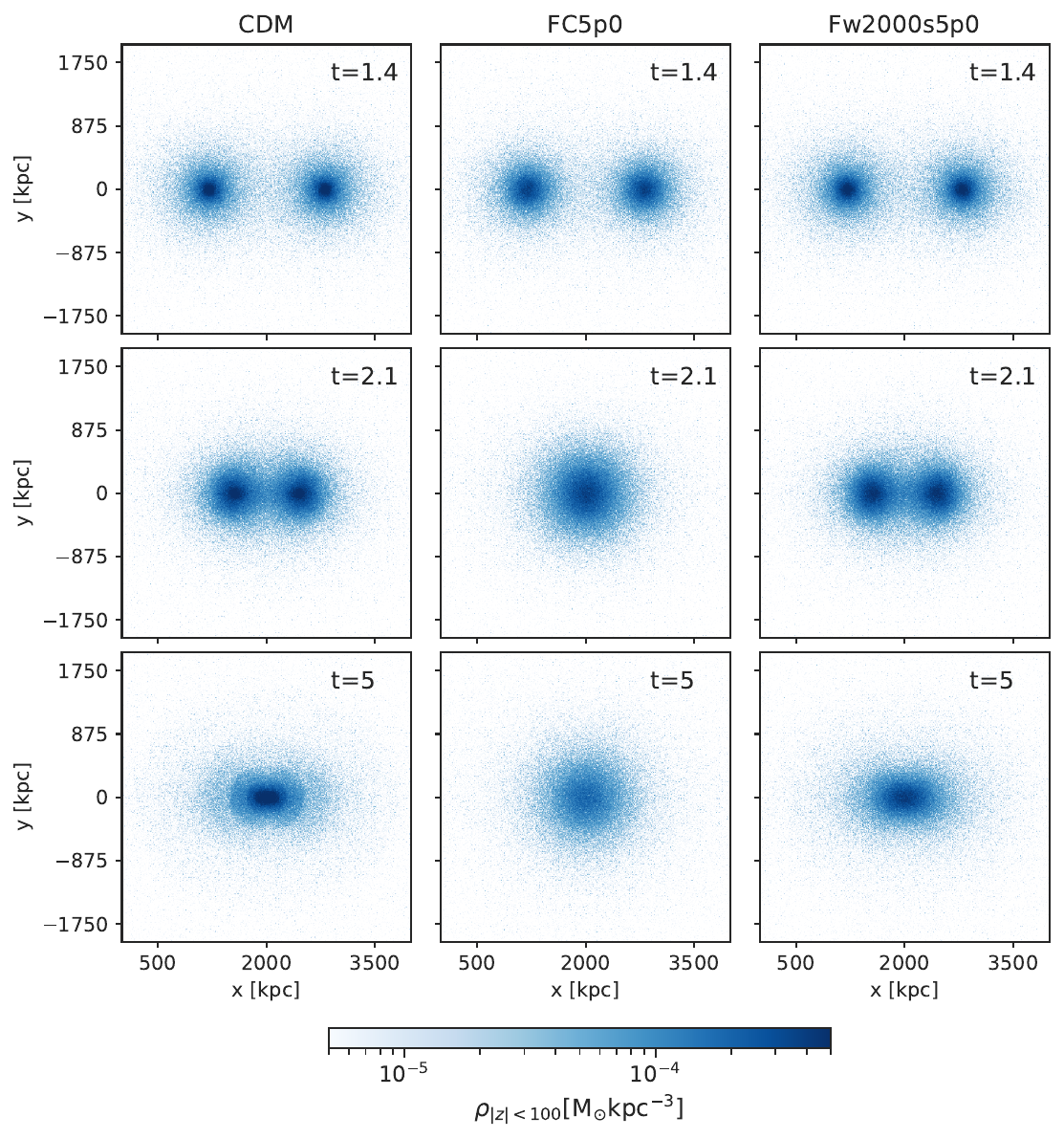}
    \caption{The density of DM haloes, accounting for the particles within $|z|<\SI{100}{kpc}$. The first column corresponds to CDM simulations, while the second and the third columns correspond to simulations with the following \review{momentum transfer} cross-sections $(\som,w): (5,\infty) , (5,2000)$. The rows correspond to different times. The first row being before the pericentre passage, the second, just after the first pericentre passage and the third being at later stages. The time stamps in the images are in units of Gyr.} 
    \label{fig:morphology s5p0 R1}
\end{figure*}

\begin{figure*}
    \centering
    \includegraphics[width=0.8\textwidth]{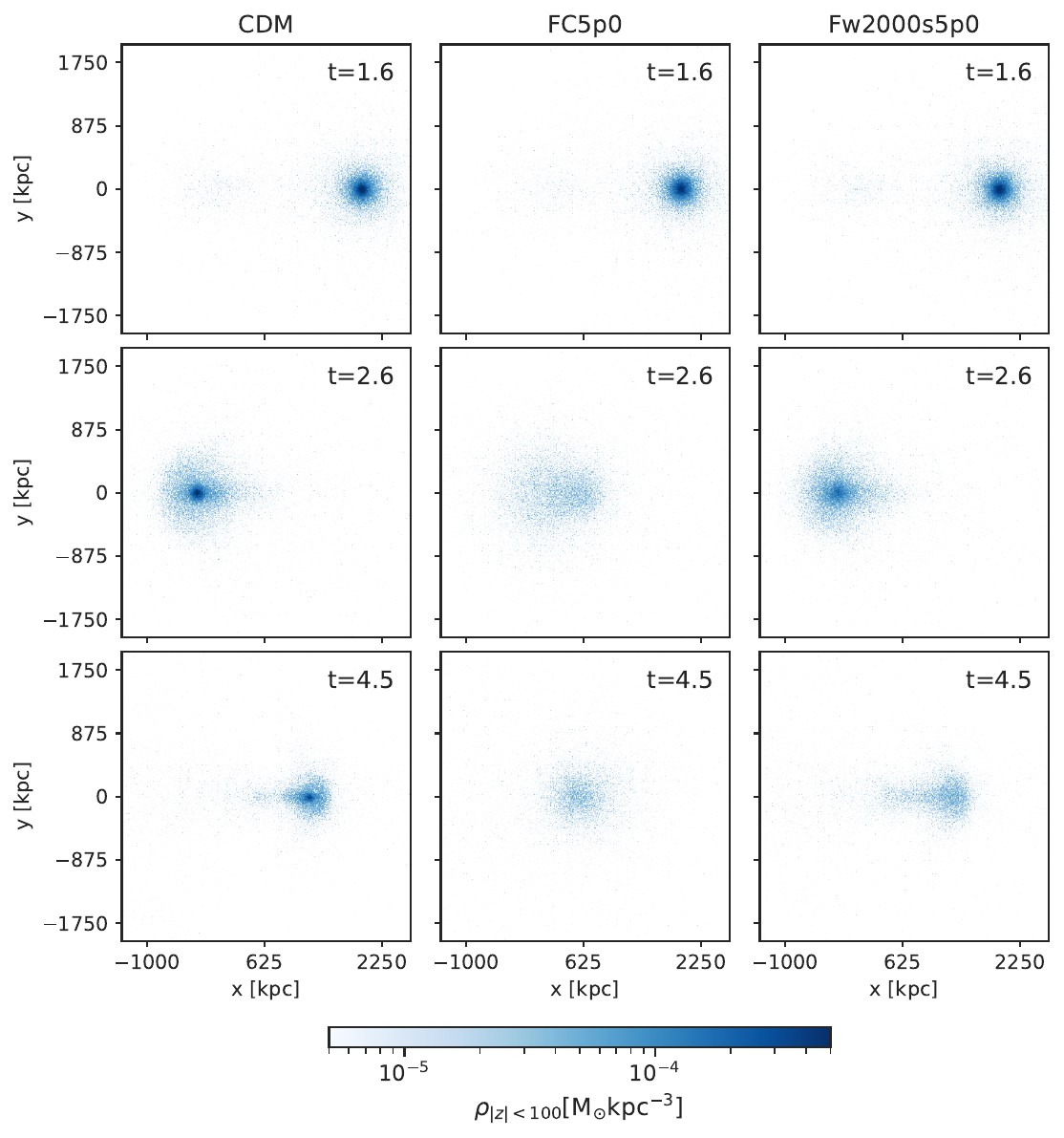}
    \caption{Plot similar to \autoref{fig:morphology s5p0 R1}, but displaying only the subhalo of MMR:5 simulation.}
    \label{fig:morphology s5p0 R5}
\end{figure*}

In \autoref{fig:morphology s5p0 R1}, we show the physical density of DM within the slice $|z| < \SI{100}{\kpc}$  projected on the merger plane. There are three columns, each correspond to a DM model and each row corresponds to a particular simulation time. First we look at time $t=\SI{1.4}{Gyr}$, i.e. before the first pericentre passage. Both haloes in the \review{constant $\sigma_T$ } simulation (column 2) have lower central densities than in the  CDM (column 1) and velocity-dependent simulations (column 3). The second row shows results at a time $t=\SI{2.1}{Gyr}$, which is just after the first pericentre passage. Owing to the large self-interaction strength, the haloes in the \review{constant $\sigma_T$ } simulation have coalesced, while for CDM and velocity-dependent self-interactions the DM haloes pass through each other. {At later stages, $t>\SI{5}{Gyr}$, the merger remnant in the \review{constant $\sigma_T$ } simulation has a lower central density. While for the velocity-dependent simulations, the effects of velocity-dependence slowly becomes relevant  as the system slows down. Thus, this leads to more cored distribution at the centre of the merger remnant when compared to the one from CDM simulation.} This feature essentially leads to the persistent BCG oscillations at late stages. \review{In \autoref{fig:morphology s5p0 R5} we have }  a similar plot displaying only the subhalo's projected density for MMR:5. Independent of the \review{choice for the $\sigma_T(v)$}, the subhaloes are observed to evaporate with time. With self-interactions, the evaporation is more pronounced. At $t=\SI{4.5}{Gyr}$, the subhalo experiencing \review{constant $\sigma_T$ } has its core dissolved significantly and comes to rest. For CDM, the core has remained relatively intact. Finally, in the velocity-dependent simulation, the core has dissolved. Although the merger remnant seems to be oscillating even at these stages. In addition, in the CDM simulation, we see shell-like features. These features are missing in the \review{constant $\sigma_T$ } simulation, since the haloes have coalesced.

\section{Central Density matched cross-section} \label{sec:results_cdens_matched}

In this section, we explore a more refined matching procedure based on simulations of isolated haloes.
In this matching scheme, parameters $\{\som, w\}$ are chosen such that different parameter combinations lead to similar central density evolution. We will refer to parameters matched according to this scheme as CD-matched. To avoid performing multiple simulations to find the matched parameter set, we make use of the self-similar nature of the gravothermal fluid equations of an isolated halo \citep{balbergSelfInteractingDark2002,essigConstrainingDissipativeDark2019}. This allows us to obtain the central density evolution without running a suite of simulations. In \cite{balbergSelfInteractingDark2002}, they assume that the \review{total} cross-section is velocity independent. In order to illustrate the rescaling, consider two constant cross-section parameters $\som^A, \som^B$. Then, the central density evolution obeys the following scaling relation:

\begin{equation}\label{eqn:rescaling central density - constant cross-section}
     \rho_c(t^A) = \rho_c\left(t^B \times \frac{\som^B}{\som^A}\right) ,
\end{equation}
where $t^A, t^B$ correspond to the evolution time of the isolated haloes simulated with parameters $\som^A,\som^B$. 

Velocity-dependent cross-sections contain two parameters, and it is not immediately clear how the central densities can be rescaled. \cite{yangGravothermalEvolutionDark2022} propose an effective cross-section $\seff$ to model the halo evolution. For a differential cross-section $\dd\sigma/\dd\cos\theta$, the effective cross-section is given by

\begin{equation}\label{eqn:effective cross-section}
    \seff = \frac{1}{512 (\sigma_{\rm 1 D})^8} \int v^2 \dd v \ \dd\cos \theta \ v^5 \sin^2 \theta \frac{ {\rm d} \sigma}{ {\rm d}
\cos \theta}\exp \left( - \frac{v^2}{4 \sigma_{\rm 1 D}^2}
\right) .
\end{equation}
In the expression given above, $v$ is the relative velocity of DM particles, $\sigma_{\rm 1D}$ is the characteristic velocity dispersion of the halo. \cite{yangGravothermalEvolutionDark2022} show that the evolution of central density in a simulation with the differential cross-section can be mimicked by a constant cross-section simulation with the same $\seff$. They also note that the equivalence holds well when the halo is in short-mean-free-path regime. In the long-mean-free-path regime, $\seff$ does not capture the effects of self-scattering accurately. However, it provides a reasonable approximation to the halo evolution. Therefore, we extend the rescaling procedure given in \eqref{eqn:rescaling central density - constant cross-section} to any differential cross-section by using the corresponding effective cross-section $\seff$.

Integrating the angular part of \eqref{eqn:effective cross-section}, we get
\begin{equation}\label{eqn:effective cross-section in terms of viscosity cross-section}
    \seff \propto \int v^2 \dd v v^5 \sigma_\mathrm{V}(v,w) \exp\left({-\frac{v^2}{4\sigma_{\rm 1D}^2}}\right) \propto \som f(w) ,
\end{equation}
where,
\begin{equation}
    \sigma_\mathrm{V} = \int \dd\cos\theta\sin^2\theta \frac{\dd\sigma}{\dd\cos\theta},
\end{equation}
is the viscosity cross-section.

Thus, for a given $w$, $\seff^A / \seff^B = \som^A / \som^B $. In other words, rescaling by $\seff$ is equivalent to rescaling by $\som$ for the same $w$. We verify this method by performing tests with some parameter combinations, see Appendix~\ref{app:central density rescaling}. 

In order to find the value $\som$ given a $w$, such that the evolution of the central density matches that of a target simulation with parameter set, $Q = \{\som^Q,w^Q\}$ we follow the procedure given below.

\begin{enumerate}
    \item Simulate an isolated halo with the target parameter set. Find the evolution of the central density $\rho_c^Q(t^Q)$ from the simulation snapshots.
    \item Simulate an isolated halo with the parameter set  $ A = \{\som^A, w \}$, followed by the estimation of the evolution of central density $\rho_{c}^A(t^A)$ from the simulation data.
    \item To obtain the evolution $\rho_c^B(t^B)$ corresponding to the parameter set $B=\{\som^B, w\}$, rescale the time axis of the simulation A, i.e.,
    \begin{equation}
        \rho_c^B(t^B) = \rho_c^A \left(t^A \times \dfrac{\som^A}{\som^B}\right) .
    \end{equation}
    \item Repeat the previous step with different values of $\som^B$ until $\rho_c^B(t^B)$ matches $\rho_c^Q(t^Q)$.
\end{enumerate}
Thus, we have obtained the CD-matched $\som^B$ for the given $w$. 

To make a guess for the value of $\som^A$ for one of our chosen values of $w$, we solve,
\begin{equation}\label{eqn:seff initial guess}
\seff(\som^A,w) = \seff^Q.    
\end{equation}
To this end, we need a value for $\soneD$. In \cite{yangGravothermalEvolutionDark2022}, they propose to use the velocity dispersion in the central region of the halo at the maximal core expansion stage for the characteristic dispersion $\sigma_{\rm 1 D}$. Using semi-analytic modelling, \cite{outmezguineUniversalGravothermalEvolution2023} show that the 1D velocity dispersion is $0.64 V_{\rm max}$ at the maximal core expansion phase. Using $V_{\rm max} \approx 1.65 r_s \sqrt{ G \rho_0}$ for our NFW parameters, we find $\sigma_{\rm 1D} \approx \qty{980}{\km\per\s}$. On the other hand, we observe $\sigma_{\rm 1D} \approx \qty{1000}{\km\per\s}$ in our simulations. Thus, we find consistency between our simulations and the semi-analytic result. 

We calculate the initial guess $\som^A$ for rare self-interactions and use the same for frequent self-interactions. For this calculation, we use the differential cross-section given in \autoref{sec:dm model}.

\begin{figure}
    \centering
    \includegraphics[width=0.48\textwidth]{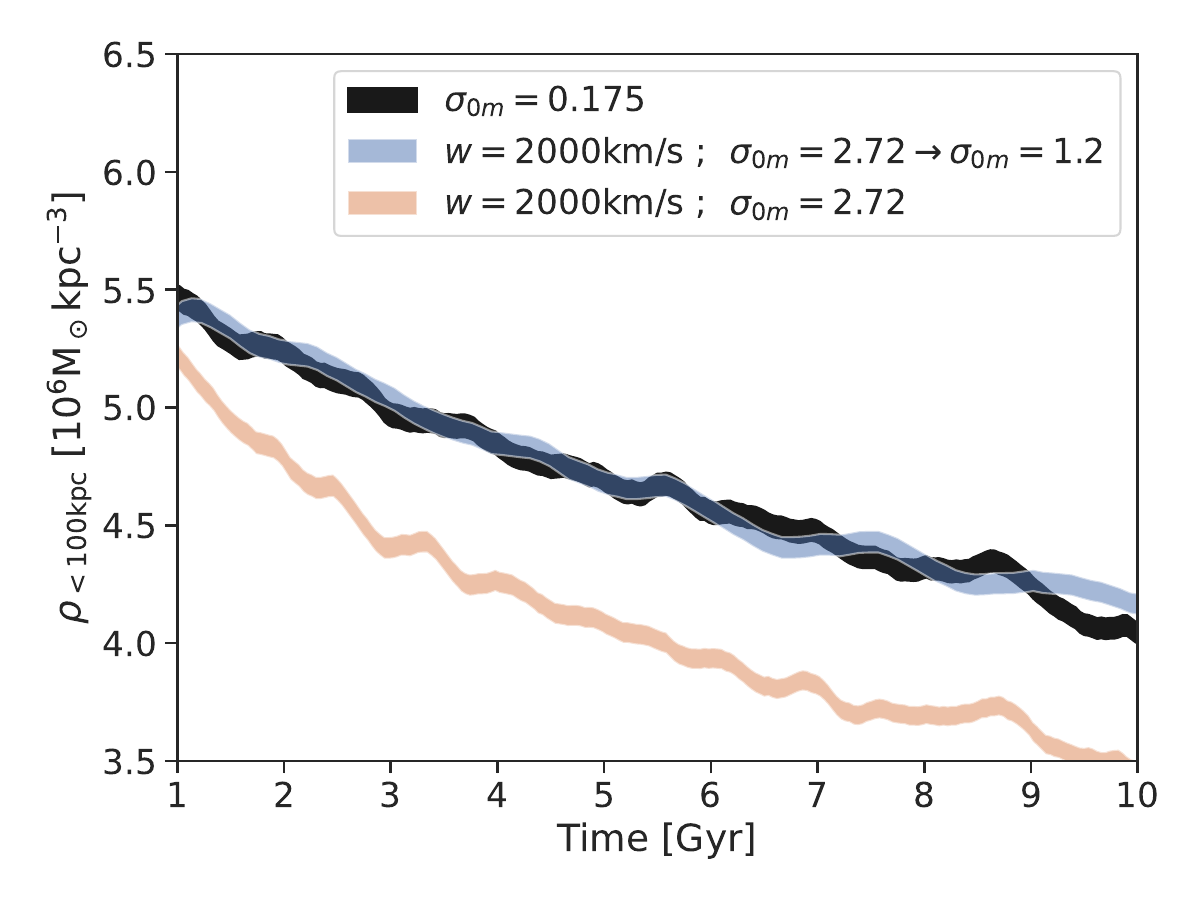}
    \caption{Evolution of central density of an isolated halo of virial mass $10^{15} \solmass$. There are three bands in the plot corresponing to, (i) simulation with rare self-interactions with \review{constant $\sigma_T$ } of $\som=0.175$ (ii) simulation for $w=\SI{2000}{\km\per\s}$ with an initial guess for $\som = \qty{2.72}{\cm\squared\per\g} $ (iii) the same simulation as the earlier one, but rescaled to $\som=\SI{1.2}{\cm\squared\per\g}$. The width of the band corresponds to the uncertainty in the estimation of central density, and it is proportional to $1/\sqrt{N}$. }
    \label{fig:cd-matching}
\end{figure}

\begin{figure}
    \centering
    \includegraphics[width=0.48\textwidth]{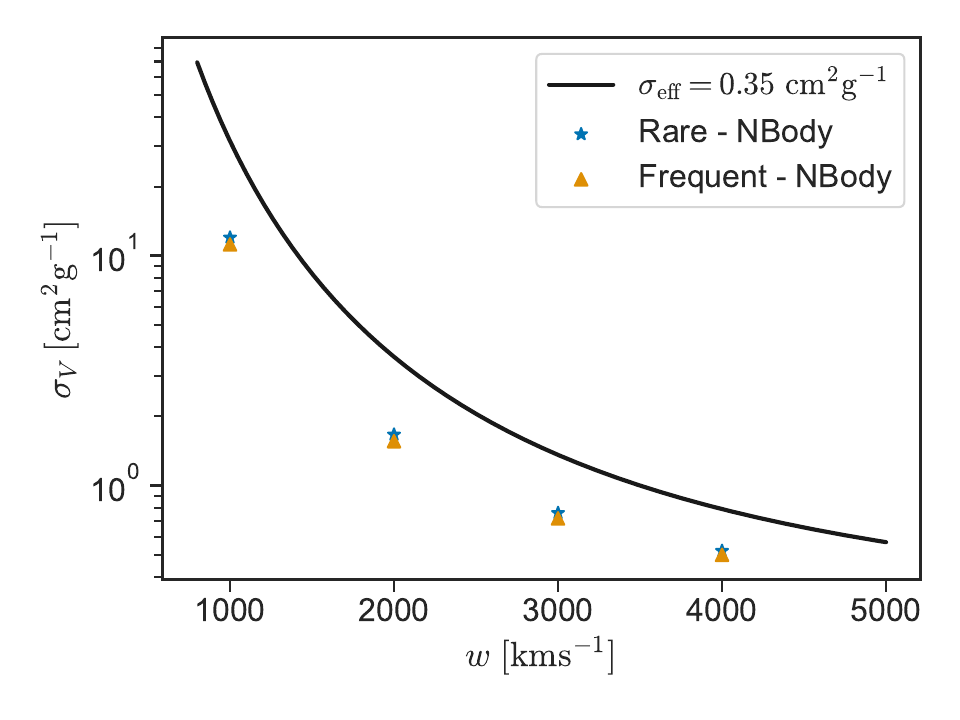}
    \caption{The plot contains viscosity cross-section evaluated at $v=0$, for different values of $w$. At each $w$, the inferred value of $\som$ is used to compute $\sigma_V$. The triangles and stars represent $\sigma_V$ calculated using the $\som$ inferred from $N$-body simulations for fSIDM and rSIDM, respectively. The solid line corresponds to $\som$ inferred by solving $\seff = \SI{0.35}{\cm\squared\per\g}$. }
    \label{fig:sigma-vs-w}
\end{figure}

We choose the target set $Q$ to correspond to the values quoted in \cite{sagunskiVelocitydependentSelfinteractingDark2021}. They quoted a 95\%  \review{upper limit} on \review{$\sigma/m_\chi$} of \SI{0.35}{\cm\squared\per\g}, assuming isotropy and velocity independence.  Therefore, this value translates to $\som=\SI{0.175}{\cm\squared\per\g}$ for rare self-interactions (because $\sigma = 2 \sigma_\mathrm{T}$). In other words, we choose the target set $Q=\lbrace\SI{0.175}{\cm\squared\per\g},\infty \rbrace$. In \autoref{fig:cd-matching}, we show an example for the matching procedure in detail. The black band corresponds to the target central density of \review{constant $\sigma_T$ } with $\som=\SI{0.175}{\cm\squared\per\g}$. The orange band corresponds to a simulation with $\lbrace \SI{2.72}{\cm\squared\per\g},\SI{2000}{\km\per\s}\rbrace$. This value for $\som$ is our initial guess calculated using \eqref{eqn:seff initial guess}. After rescaling by trial and error, the desired value of $\som$ is found to be $\SI{1.2}{\cm\squared\per\g}$. This procedure can be extended to all chosen values of $w$ and the obtained results are tabulated in \autoref{tab:upper bound params}. The inferred values of $\som$ at different values of $w$ can then be used to calculate the corresponding viscosity cross-section $\sigma_V$. This is shown in \autoref{fig:sigma-vs-w}. The orange triangles and blue stars represent the values of $\sigma_V$ obtained using the inferred values of $\som$ from $N$-body simulations for rare and frequent self-interactions, respectively. Similarly, the solid line corresponds to the $\sigma_V$ calculated from the values of $\som$ inferred by solving \eqref{eqn:seff initial guess}. The fact that the results obtained from $\seff$ \review{and} $N$-body simulations are different can be attributed to the fact that the isolated halo is in the long-mean-free-path regime. This was already noted in \cite{yangGravothermalEvolutionDark2022}.

The evolution of an isolated halo has a feature that, as long as it is in the core expansion phase, at any given time the central density is monotonically decreasing with $\seff$. This implies that for a given value of $w$,  and at a given time in the evolution of the halo, core-size is larger for larger values of $\som$. \cite{sagunskiVelocitydependentSelfinteractingDark2021},\cite{andradeStringentUpperLimit2021},and \cite{eckertConstraintsDarkMatter2022a} constrain $\som$ using the observed core-size in clusters. Hence, for every value of $w$, there is a value of $\som$ that produces core sizes, or central densities, similar to the current upper bound. Increasing $\som$ any further would increase the core size to values larger than what is observed. Thus, this matching procedure can be used to extend the bounds from \review{constant $\sigma_T$ } to different values of $w$.

When matched using $\seff$ or central density evolution, we observe that the ratio between the $\som$'s of rare and frequent is approximately 0.64 at every chosen values of $w$. This can be understood from the definition of $\seff$. From \eqref{eqn:effective cross-section in terms of viscosity cross-section}, we have for any $w$, for rare self-interactions,

\begin{align}
    \seff &= C \som \int \dd \theta \sin^3\theta \int \dd v v^7 \exp\left(\frac{-v^2}{4 \soneD^2}\right) \left(1+\frac{v^2}{w^2}\right)^{-2} \\
          &= C \frac{4}{3} \som f(w) .
\end{align}
Here, $C = 1/(512 \soneD^8)$. Now from \eqref{eqn:momentum transfer cross-section used in sims}, for rare self-interactions we have $\sigma_\mathrm{T} = \som g(w,v)$, which implies that

\begin{equation}\label{eqn:rare sigma effective and momentum transfer}
    \seff= \frac{4}{3} \sigma_\mathrm{T} Cf/g .
\end{equation}
For frequent self-interactions, we consider a differential cross-section of the form given in \eqref{eqn:differential cross-section + generic form} with the function $\Theta(\theta)$ having support only for values of $\theta$ close to 0. A simple choice for $\Theta(\theta)$ is a step-function that is non-zero in the interval $[0,\epsilon]$, where $\epsilon$ is some small number. Therefore, $\seff$ is given as

\begin{align}
    \seff &= C N  \som \int \dd \theta \sin^3\theta \Theta(\theta) \int \dd v v^7 \exp\left(\frac{-v^2}{4 \soneD^2}\right) g(w,v) \\
          &\approx C N \som \int \dd \theta \theta^3 \Theta(\theta) f(w) = C N \som f(w) \epsilon^4/4,
\end{align}
where in the second equality we have Taylor-expanded and retained only the leading order in $\theta$ in the angular integrand.
Similarly, for the momentum transfer cross-section we have

\begin{align}
    \sigma_\mathrm{T} &= N \som g(w,v) \int \dd \theta \sin\theta (1 - |\cos\theta|) \Theta(\theta) \\
    &\approx N \som g(w,v) \epsilon^4/8 .
\end{align}
Thus for frequent self-interactions, we have
\begin{equation}
\label{eqn:freq sigma effective and momentum transfer}
    \seff = 2 \sigma_\mathrm{T} C f / g .
\end{equation}
Now for matching with $\seff$ or the central density we require $\seff({\rm Rare}) = \seff({\rm Freq.})$.  Upon using \eqref{eqn:rare sigma effective and momentum transfer} and \eqref{eqn:freq sigma effective and momentum transfer}, this requirement leads to the matching condition \review{$\sigma_\mathrm{T}({\rm Freq.}) = (2/3) \sigma_\mathrm{T}({\rm Rare}) \implies \som({\rm Freq.}) = (2/3)\som({\rm Rare}) $} as seen in \autoref{tab:upper bound params}. 

\subsection{Central density matched simulations – qualitative features}

In this subsection, we look at the qualitative differences in mergers when parameters are chosen according to the central density matching procedure.  Again, as in \autoref{sec:varying w}, we only simulate frequent self-interactions. \reviewA{In \autoref{sec:varying w}, we investigated the effects arising from changing the value of $w$. To ensure that the effects of self-interaction are not negligible for $w=\SI{1000}{\km\per\s}$, we used a large value for $\som$ of \SI{5.0}{\cm\squared\per\g}.  On the other hand, in this section the cross-section parameters are CD-matched for which we choose the target set to be $Q=\{\SI{1.0}{\cm\squared\per\g} , \infty\}$. The matched parameters are given in \autoref{tab:seff 1.5 labels}. We have chosen a value for $\som $ such that the effects of self-interaction are observable, but not as large as the previous value \SI{5.0}{\cm\squared\per\g}.}

\begin{figure}
    \centering
    \includegraphics[width=0.48\textwidth]{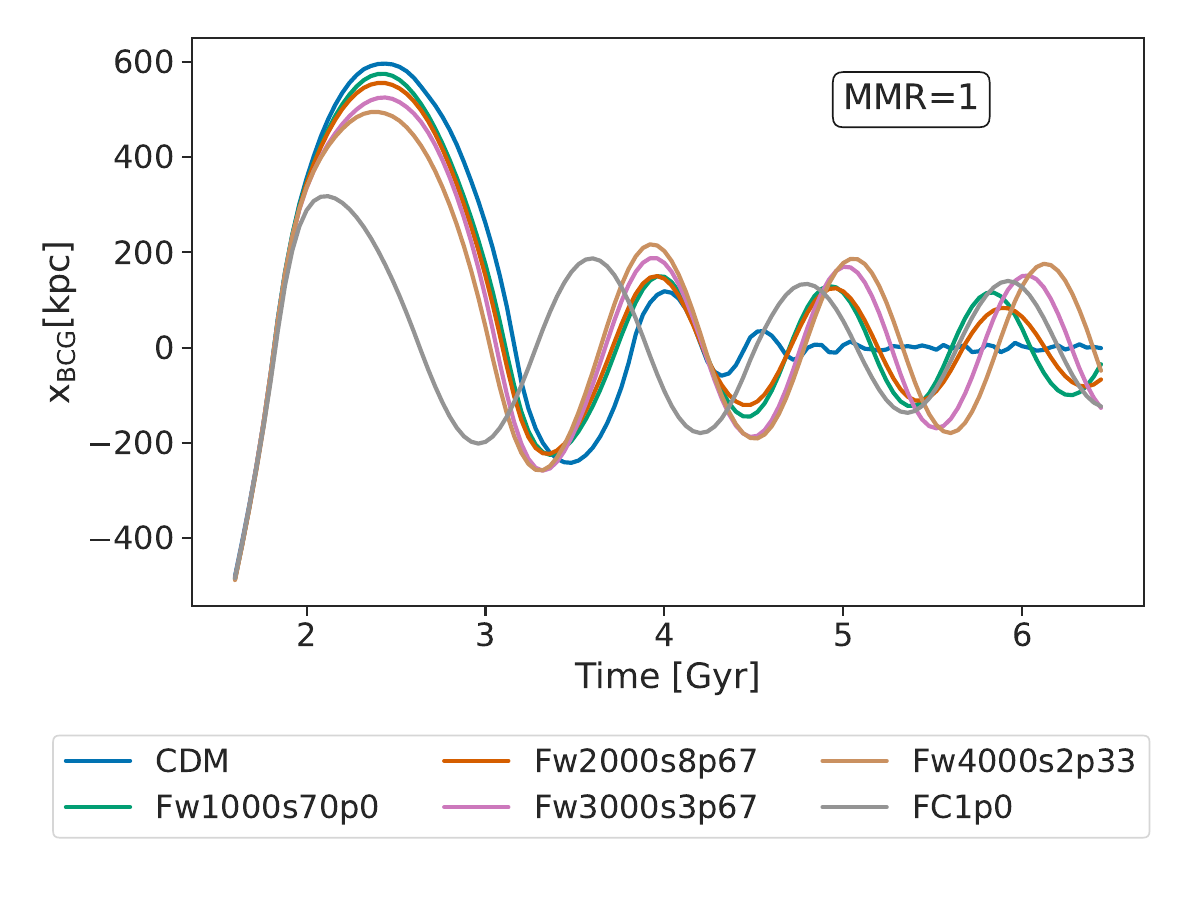}
    \includegraphics[width=0.48\textwidth]{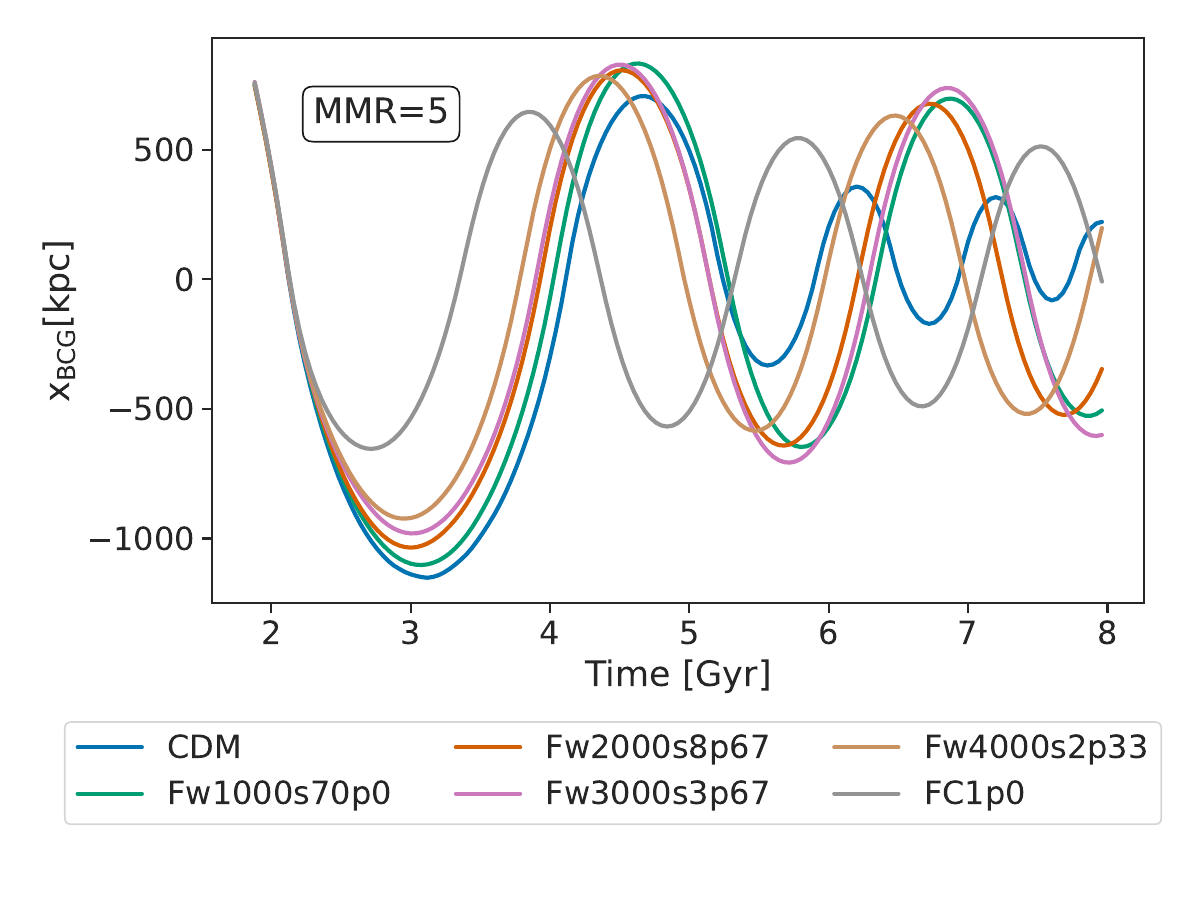}
    \caption{BCG positions of the subhalo against time. The upper panel corresponds to equal mass merger, while the lower one corresponds to unequal mass merger.}
    \label{fig:BCG peaks seff}
\end{figure}

For brevity, we show only the BCG oscillations in \autoref{fig:BCG peaks seff}. We see that the BCG oscillations of velocity-dependent simulations all have similar amplitudes at early stages. Only at a later stages they start to deviate. This similarity stems from the fact that the parameters are CD-matched.  Similar to what was explained in \autoref{sec: bcg oscillations}, at early times the DM particles have a large relative velocity. As a result, most of the velocity-dependent cross-sections have a small effective self-interaction strength. At later stages, the system slows down and $v \lesssim w$ and the effective self-interaction strength increases. In addition, the internal evolution around a DM peak is similar for all the cross-section parameters chosen, since they are CD-matched. See appendix \ref{app:cd evolution in merger} for the evolution of the central density of the main halo.

\begin{table}
\centering
\renewcommand{\arraystretch}{1.5}
\begin{tabular}{ccc}
\hline
\multicolumn{1}{c|}{$w$}                     & \multicolumn{1}{c|}{$\som$}                       & \multirow{2}{*}{Label} \\
\multicolumn{1}{c|}{$[\SI{}{\km\per\s}]$} & \multicolumn{1}{c|}{$[\SI{}{\cm\squared\per\g}]$} &                        \\ \hline
\multicolumn{1}{c|}{1000}     & \multicolumn{1}{c|}{70.0} & Fw1000s70p0 \\
\multicolumn{1}{c|}{2000}     & \multicolumn{1}{c|}{8.67} & Fw2000s8p67 \\
\multicolumn{1}{c|}{3000}     & \multicolumn{1}{c|}{3.67} & Fw3000s3p67 \\
\multicolumn{1}{c|}{4000}     & \multicolumn{1}{c|}{2.33} & Fw4000s2p33 \\
\multicolumn{1}{c|}{$\infty$} & \multicolumn{1}{c|}{1.0}  & FC1p0       \\ \hline
\end{tabular}
\caption{This table contains labels and cross-section parameters matched to constant cross-section, $\som=\SI{1.0}{\cm\squared\per\g}$ of frequent self-interactions using central density matching scheme.   }\label{tab:seff 1.5 labels}
\end{table}

\subsection{Central density matched upper bound simulations}

In \review{\autoref{sec:results_cdens_matched}}, the conservative upper bounds were obtained using the central density matching scheme. Values are tabulated in \autoref{tab:upper bound params}. We run merger simulations for this set of parameters and estimate the offsets $d_{\rm DM-BCG}$. For velocity-independent cross-sections up to $\SI{0.5}{\cm\squared\per\g}$, \cite{fischerNbodySimulationsDark2021a} found that the DM-BCG and DM-Galaxy offsets increase with increasing values of $\som$. By running merger simulations at the upper bound values of the cross-section parameters, we estimate the order of magnitude of the largest possible offsets allowed by current bounds.

\begin{table*}
\centering
\renewcommand{\arraystretch}{1.5} 
\begin{tabular}{c|cc|cc}
$w$      & \multicolumn{2}{c|}{Frequent}             & \multicolumn{2}{c}{Rare}                  \\ \cline{2-5} 
         & \multicolumn{1}{c|}{$\som$} & Label       & \multicolumn{1}{c|}{$\som$} & Label       \\
$[\SI{}{\km\per\s}]$ & \multicolumn{1}{c|}{$[\SI{}{\cm\squared\per\g}]$} &  & \multicolumn{1}{c|}{$[\SI{}{\cm\squared\per\g}]$} &  \\ \hline
1000     & \multicolumn{1}{c|}{5.6}    & Fw1000s5p6  & \multicolumn{1}{c|}{9.0}    & Rw1000s9p0  \\
2000     & \multicolumn{1}{c|}{0.78}   & Fw2000s0p78 & \multicolumn{1}{c|}{1.25}   & Rw2000s1p25 \\
3000     & \multicolumn{1}{c|}{0.36}   & Fw3000s0p36 & \multicolumn{1}{c|}{0.57}   & Rw3000s0p57 \\
4000     & \multicolumn{1}{c|}{0.25}   & Fw4000s0p25 & \multicolumn{1}{c|}{0.39}   & Rw4000s0p39 \\
$\infty$ & \multicolumn{1}{c|}{0.11}   & FC0p11      & \multicolumn{1}{c|}{0.175}  & RC0p175    
\end{tabular}
\caption{This table contains the conservative upper bound on $\som$ for different values of $w$ given in the first column. The second column contains $\som$ the value for frequent self-interactions, and the third for rare self-interactions.}\label{tab:upper bound params}
\end{table*}

\begin{figure}
    \centering
    \includegraphics[width=0.48\textwidth]{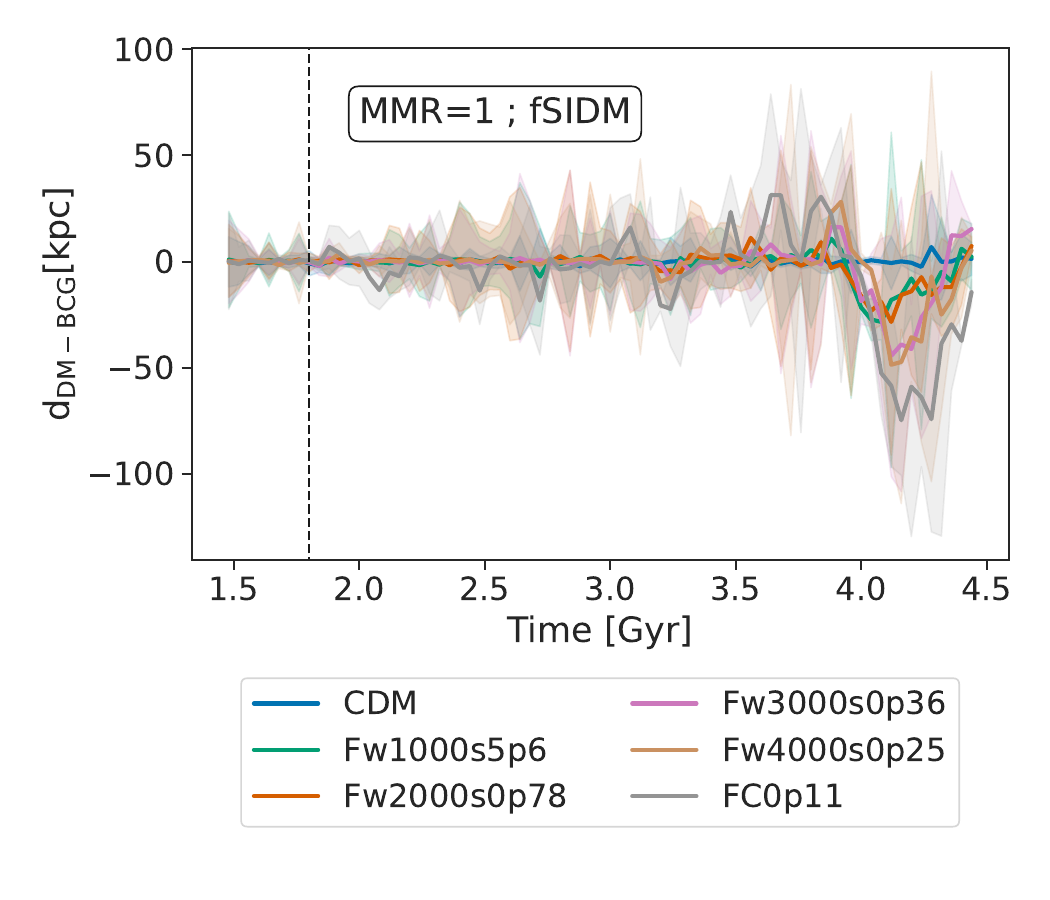}
    \includegraphics[width=0.48\textwidth]{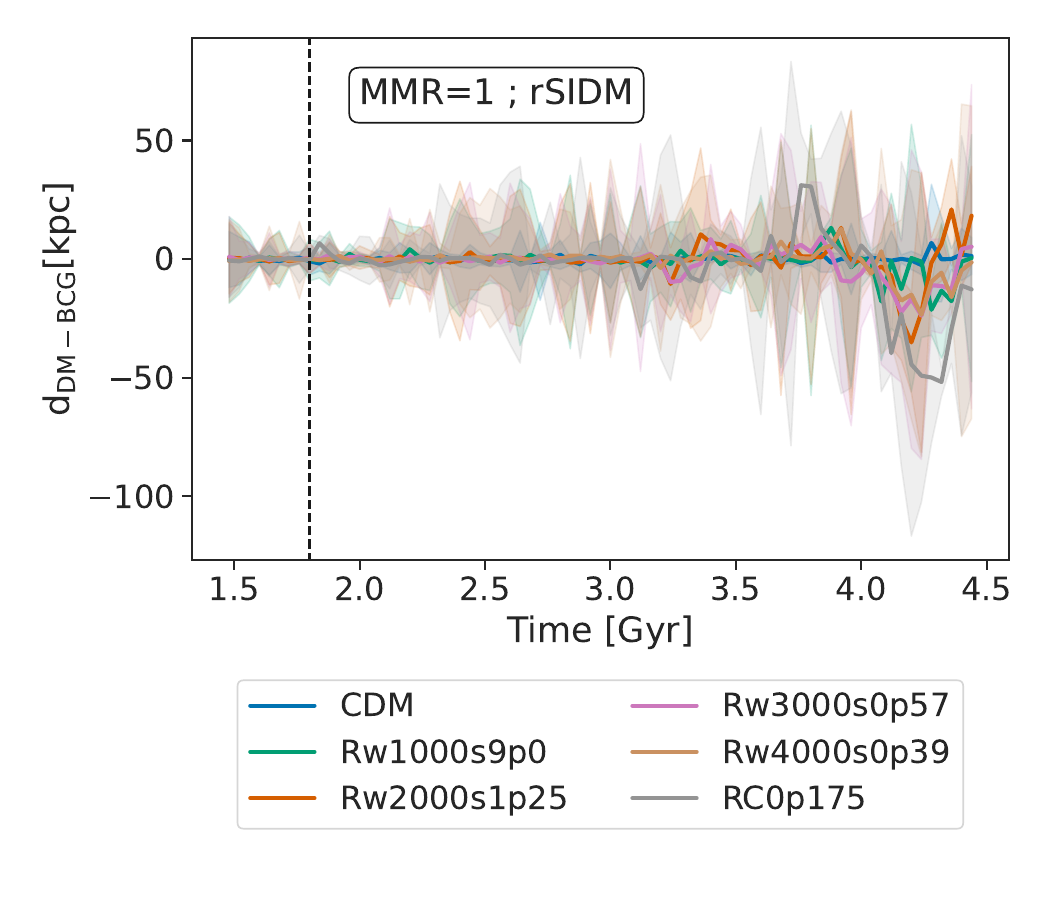}
    \caption{DM – BCG offsets in the equal-mass merger. Upper panel corresponds to fSIDM, while bottom panel to rSIDM. The plot labels are described in \autoref{tab:upper bound params}. The vertical dashed lines correspond to the time of the first pericentre passage.
    }
    \label{fig:upper bound dm bcg offset}
\end{figure}
In \autoref{fig:upper bound dm bcg offset}, we show the DM-BCG offset at the first pericentre passage for the equal mass merger. The offset is $\mathcal{O}(1) \kpc$, while the offsets after the third pericentre passage ($t \sim 4$ billion years) start increasing \review{and is seen to be $\mathcal{O}(10) \kpc$}. The effective self-interaction strength is not large enough to produce an observable offset at the first pericentre. Hence,  it might be difficult to find such an offset in real observations. We do not show the offsets for unequal mass merger because, the offsets just after the first pericentre are smaller than the ones of equal mass merger. In addition, at later stages due to the evaporation of halo, we do not have reliable peak positions. We can also see from \autoref{fig:upper bound dm bcg offset} that the offsets produced by fSIDM are larger than that of rSIDM, and this is due to the fact that mergers are sensitive to the angular dependence.

\section{Conclusions} \label{sec:conclusion}

We first discuss the assumptions made in the paper before we conclude. The first assumption that we make is that we use idealized initial conditions.  \review{Yet}, it is informative to study them to find the appropriate features to look out for in more realistic simulations and observations. One example is the amplitude of BCG oscillation at late stages, which is seen to depend on the velocity-dependent cross-section parameters. Therefore, it is instructive to simulate cosmological boxes with velocity-dependent cross-section at higher resolution in order to estimate the distribution of DM-BCG offsets. Later, this could be compared to observations \citep[e.g.][]{crossExaminingSelfInteractionDark2023,lauerBrightestClusterGalaxies2014}. For example, \cite{harveyObservableTestsSelfinteracting2019} studied the DM-BCG offsets in the BAHAMAS–SIDM suite of cosmological simulations and placed constraints on $\sigma/m_\chi$ assuming velocity-independent isotropic SIDM.  

In addition, we have modelled the BCG and galaxies as collision-less point particles. \reviewA{A more realistic treatment of BCG's and galaxies is necessary for better comparison with observations.}
Furthermore, we neglect the effect of galaxies having their own DM halo \citep{kummerEffectiveDescriptionDark2018}.
In reality, approximately 10\% of the mass of clusters is made up by the intracluster medium (ICM), which is not included in our simulations. Merger studies that include the ICM, such as \cite{robertsonWhatDoesBullet2017}, find that the DM-galaxy offset is not significantly affected by the presence of the ICM. \cite{fischerRoleBaryonsSelfinteracting2023} also finds similar results. Therefore, we argue that the absence of an ICM component does not significantly affect our conclusions.\review{ We leave the study of these systems with hydrodynamic simulations to a future study.}

As mentioned in \autoref{sec:dm model}, we have assumed that the angular and velocity dependence of the differential cross-sections can be separated into two functions. This assumption has to be dropped when dealing with realistic SIDM models \citep{tulincollisionlessDarkMatter2013,fengHiddenChargedDark2009}. There are other SIDM models leading to different effects that are not included in our studies. For example, in addition to elastic scattering, inelastic scattering could be included \citep{oneilEndothermicSelfinteractingDark2023}. We leave the study of such models to future work. 

 On the observation side, all observed DM-BCG offsets are inferred to be just after pericentre passage and have uncertainties that make them consistent with zero \citep{bradacRevealingPropertiesDark2008,dawsonDiscoveryDissociativeGalaxy2012,dawsonConstrainingDarkMatter2013,jeeWeighingGordoPrecision2014,jeeMCConstrainingDark2015}. Similarly, estimating DM-galaxy offsets are difficult due to shot noise arising from the smaller galaxy count. In our simulations we had $10^6$ galaxy particles, but in reality we observe at most $\sim 10^3$ of them. 
 On the other hand, \cite{lauerBrightestClusterGalaxies2014} find a median offset of $\sim \SI{10}{\kpc}$, with the offset measured between BCG and cluster centre, the latter being identified by X-ray observations. Their sample comprised 433 BCGs that are located in Abell galaxy clusters. The DM-BCG offset distribution could be compared to predictions of cosmological simulations. Overall, the situation with observations is expected to improve with forthcoming surveys, such as \textit{SuperBIT} \citep{romualdezDesignDevelopmentHighresolution2016} and \textit{Euclid} \citep{laureijsEuclidDefinitionStudy2011a}.

We have simulated idealized, isolated haloes and galaxy cluster mergers with equal and unequal merger mass ratios with velocity-dependent, frequent and rare self-interactions. Mergers are interesting astrophysical probes since the system is sensitive to self-interaction cross-sections with both angular and velocity dependencies. Therefore, we focused on understanding the qualitative effects that arise from velocity dependence in mergers. On the quantitative side, we also investigate the maximum offsets that can be observed given the current bounds on $\sigma/m_\chi$. 

\begin{itemize}
    \item Independent of the matching procedure used in the paper, the effects of velocity-dependent cross-sections can be observed on galaxy cluster mergers by comparing the early time and late time oscillations of BCG. In particular, the degeneracy in the cross-section parameters when studying the evolution of central density in isolated haloes is broken when studying mergers. This is due to the fact that the relative velocities of the merging clusters change with time. 
    \item The evolution of central densities of isolated haloes are similar between rare and frequent self-interaction, when the momentum transfer cross-section $\sigma_\mathrm{T}(v)$ of fSIDM is chosen to be $2/3 \sigma_\mathrm{T}(v)$ of rSIDM. The factor 2/3 follows from matching the angular dependence of fSIDM and rSIDM with viscosity cross-section, as seen in \autoref{fig:sigma-vs-w}.
    \item We extend the existing upper bounds on the constant cross-section $\som$ to the parameter space $\lbrace\som,w\rbrace$ of velocity-dependent, rare and frequent self-interactions. 
    \item In the equal-mass merger simulations with upper-bound cross-section parameters, we find that the offsets after the first pericentre is approximately $\mathcal{O}(1) \SI{}{kpc}$. In particular, the offsets are the largest in the constant cross-section simulation. As the system evolves further, offsets grow. After the third pericentre passage, due to the oscillations of BCG, and the galactic component the offsets are $\mathcal{O}(10) \SI{}{kpc}$. Thus, mergers in their late stages are interesting to test and constrain SIDM.
\end{itemize}

\review{In conclusion, we have studied the qualitative effects of velocity-dependent SIDM cross-sections in galaxy cluster mergers. Our models do not have the realism required for a direct comparison with astronomical data, owing to the neglection of baryonic effects.  However, they offer insights into the physical processes that govern the phenomenology of SIDM. More realistic predictions can be obtained by performing full hydrodynamical simulations that include stars, cooling and feedback effects. The significantly larger complexity of such models with additional degrees of freedom render the interpretation much harder. Clearly, this is an avenue for future work.}

\section*{Acknowledgements}

\reviewA{We want to thank the anonymous referee for helpful comments that improved the paper. We would also like to} thank all participants of the \href{https://darkium.org/#}{Darkium} SIDM Journal Club for helpful discussions. 
This work is funded by the Deutsche Forschungsgemeinschaft (DFG, German Research Foundation) under Germany's Excellence Strategy -- EXC 2121 ``Quantum Universe'' --  390833306, Germany’s Excellence Strategy -- EXC-2094 ``Origins'' -- 390783311, and the Emmy Noether Grant No.\ KA 4662/1-2.
Preprint numbers: DESY-23-153, TTP23-043.
\section*{Data Availability}

The data underlying this article will be shared on reasonable request to the corresponding author.
 



\bibliographystyle{mnras}
\bibliography{velDepMergers,unpublished} 

\begin{thebibliography}{}
\makeatletter
\relax
\def\mn@urlcharsother{\let\do\@makeother \do\$\do\&\do\#\do\^\do\_\do\%\do\~}
\def\mn@doi{\begingroup\mn@urlcharsother \@ifnextchar [ {\mn@doi@}
  {\mn@doi@[]}}
\def\mn@doi@[#1]#2{\def\@tempa{#1}\ifx\@tempa\@empty \href
  {http://dx.doi.org/#2} {doi:#2}\else \href {http://dx.doi.org/#2} {#1}\fi
  \endgroup}
\def\mn@eprint#1#2{\mn@eprint@#1:#2::\@nil}
\def\mn@eprint@arXiv#1{\href {http://arxiv.org/abs/#1} {{\tt arXiv:#1}}}
\def\mn@eprint@dblp#1{\href {http://dblp.uni-trier.de/rec/bibtex/#1.xml}
  {dblp:#1}}
\def\mn@eprint@#1:#2:#3:#4\@nil{\def\@tempa {#1}\def\@tempb {#2}\def\@tempc
  {#3}\ifx \@tempc \@empty \let \@tempc \@tempb \let \@tempb \@tempa \fi \ifx
  \@tempb \@empty \def\@tempb {arXiv}\fi \@ifundefined
  {mn@eprint@\@tempb}{\@tempb:\@tempc}{\expandafter \expandafter \csname
  mn@eprint@\@tempb\endcsname \expandafter{\@tempc}}}

\bibitem[\protect\citeauthoryear{Ackerman, Buckley, Carroll  \&
  Kamionkowski}{Ackerman et~al.}{2009}]{ackermanDarkMatterDark2009a}
Ackerman L.,  Buckley M.~R.,  Carroll S.~M.,   Kamionkowski M.,  2009, \mn@doi
  [PhRvD] {10.1103/PhysRevD.79.023519}, \href
  {https://ui.adsabs.harvard.edu/abs/2009PhRvD..79b3519A} {79, 023519}

\bibitem[\protect\citeauthoryear{Adhikari et~al.,}{Adhikari
  et~al.}{2022}]{adhikariAstrophysicalTestsDark2022}
Adhikari S.,  et~al., 2022, Astrophysical {{Tests}} of {{Dark Matter
  Self-Interactions}} (\mn@eprint {} {2207.10638})

\bibitem[\protect\citeauthoryear{Andrade, Fuson, {Gad-Nasr}, Kong, Minor,
  Roberts  \& Kaplinghat}{Andrade
  et~al.}{2021}]{andradeStringentUpperLimit2021}
Andrade K.~E.,  Fuson J.,  {Gad-Nasr} S.,  Kong D.,  Minor Q.,  Roberts M.~G.,
   Kaplinghat M.,  2021, \mn@doi [MNRAS] {10.1093/mnras/stab3241}, \href
  {https://ui.adsabs.harvard.edu/abs/2022MNRAS.510...54A} {510, 54}

\bibitem[\protect\citeauthoryear{Balberg, Shapiro  \& Inagaki}{Balberg
  et~al.}{2002}]{balbergSelfInteractingDark2002}
Balberg S.,  Shapiro S.~L.,   Inagaki S.,  2002, \mn@doi [ApJ]
  {10.1086/339038}, \href
  {https://ui.adsabs.harvard.edu/abs/2002ApJ...568..475B} {568, 475}

\bibitem[\protect\citeauthoryear{Binney \& Tremaine}{Binney \&
  Tremaine}{2008}]{binneyGalacticDynamicsSecond2008}
Binney J.,  Tremaine S.,  2008,
  \href{https://ui.adsabs.harvard.edu/abs/2008gady.book.....B}{Galactic
  {{Dynamics}}: {{Second Edition}}}.
{Princeton University Press}

\bibitem[\protect\citeauthoryear{Boddy, Feng, Kaplinghat  \& Tait}{Boddy
  et~al.}{2014}]{boddySelfinteractingDarkMatter2014}
Boddy K.~K.,  Feng J.~L.,  Kaplinghat M.,   Tait T. M.~P.,  2014, \mn@doi
  [PhRvD] {10.1103/PhysRevD.89.115017}, \href
  {https://ui.adsabs.harvard.edu/abs/2014PhRvD..89k5017B} {89, 115017}

\bibitem[\protect\citeauthoryear{Brada{\v c}, Allen, Treu, Ebeling, Massey,
  Morris, {von der Linden}  \& Applegate}{Brada{\v c}
  et~al.}{2008}]{bradacRevealingPropertiesDark2008}
Brada{\v c} M.,  Allen S.~W.,  Treu T.,  Ebeling H.,  Massey R.,  Morris R.~G.,
   {von der Linden} A.,   Applegate D.,  2008, \mn@doi [ApJ] {10.1086/591246},
  \href {https://ui.adsabs.harvard.edu/abs/2008ApJ...687..959B} {687, 959}

\bibitem[\protect\citeauthoryear{Bullock \& {Boylan-Kolchin}}{Bullock \&
  {Boylan-Kolchin}}{2017}]{bullockSmallScaleChallengesLambda2017}
Bullock J.~S.,  {Boylan-Kolchin} M.,  2017, \mn@doi [ARA&A]
  {10.1146/annurev-astro-091916-055313}, \href
  {https://ui.adsabs.harvard.edu/abs/2017ARA%26A..55..343B} {55, 343}

\bibitem[\protect\citeauthoryear{Cline, Liu, Moore  \& Xue}{Cline
  et~al.}{2014}]{clineScatteringPropertiesDark2014}
Cline J.~M.,  Liu Z.,  Moore G.~D.,   Xue W.,  2014, \mn@doi [PhRvD]
  {10.1103/PhysRevD.89.043514}, \href
  {https://ui.adsabs.harvard.edu/abs/2014PhRvD..89d3514C} {89, 043514}

\bibitem[\protect\citeauthoryear{Correa}{Correa}{2021}]{correaConstrainingVelocitydependentSelfinteracting2021}
Correa C.~A.,  2021, \mn@doi [MNRAS] {10.1093/mnras/stab506}, \href
  {https://ui.adsabs.harvard.edu/abs/2021MNRAS.503..920C} {503, 920}

\bibitem[\protect\citeauthoryear{Cross et~al.,}{Cross
  et~al.}{2023}]{crossExaminingSelfInteractionDark2023}
Cross D.,  et~al., 2023, Examining the {{Self-Interaction}} of {{Dark Matter}}
  through {{Central Cluster Galaxy Offsets}} (\mn@eprint {} {2304.10128})

\bibitem[\protect\citeauthoryear{Dawson}{Dawson}{2013}]{dawsonConstrainingDarkMatter2013}
Dawson W.~A.,  2013, PhD thesis, University of California, Davis, \url
  {https://ui.adsabs.harvard.edu/abs/2013PhDT.......211D}

\bibitem[\protect\citeauthoryear{Dawson et~al.,}{Dawson
  et~al.}{2012}]{dawsonDiscoveryDissociativeGalaxy2012}
Dawson W.~A.,  et~al., 2012, \mn@doi [ApJ] {10.1088/2041-8205/747/2/L42}, \href
  {https://ui.adsabs.harvard.edu/abs/2012ApJ...747L..42D} {747, L42}

\bibitem[\protect\citeauthoryear{Dutton \& Macci{\`o}}{Dutton \&
  Macci{\`o}}{2014}]{duttonColdDarkMatter2014a}
Dutton A.~A.,  Macci{\`o} A.~V.,  2014, \mn@doi [MNRAS] {10.1093/mnras/stu742},
  \href {https://ui.adsabs.harvard.edu/abs/2014MNRAS.441.3359D} {441, 3359}

\bibitem[\protect\citeauthoryear{Eckert, Ettori, Robertson, Massey,
  Pointecouteau, Harvey  \& McCarthy}{Eckert
  et~al.}{2022}]{eckertConstraintsDarkMatter2022a}
Eckert D.,  Ettori S.,  Robertson A.,  Massey R.,  Pointecouteau E.,  Harvey
  D.,   McCarthy I.~G.,  2022, \mn@doi [A&A] {10.1051/0004-6361/202243205},
  \href {https://ui.adsabs.harvard.edu/abs/2022A&A...666A..41E} {666, A41}

\bibitem[\protect\citeauthoryear{Elbert, Bullock, {Garrison-Kimmel}, Rocha,
  O{\~n}orbe  \& Peter}{Elbert et~al.}{2015}]{elbertCoreFormationDwarf2015}
Elbert O.~D.,  Bullock J.~S.,  {Garrison-Kimmel} S.,  Rocha M.,  O{\~n}orbe J.,
    Peter A. H.~G.,  2015, \mn@doi [MNRAS] {10.1093/mnras/stv1470}, \href
  {https://ui.adsabs.harvard.edu/abs/2015MNRAS.453...29E} {453, 29}

\bibitem[\protect\citeauthoryear{Essig, McDermott, Yu  \& Zhong}{Essig
  et~al.}{2019}]{essigConstrainingDissipativeDark2019}
Essig R.,  McDermott S.~D.,  Yu H.-B.,   Zhong Y.-M.,  2019, \mn@doi [PhRvL]
  {10.1103/PhysRevLett.123.121102}, \href
  {https://ui.adsabs.harvard.edu/abs/2019PhRvL.123l1102E/abstract} {123,
  121102}

\bibitem[\protect\citeauthoryear{Feng, Kaplinghat, Tu  \& Yu}{Feng
  et~al.}{2009}]{fengHiddenChargedDark2009}
Feng J.~L.,  Kaplinghat M.,  Tu H.,   Yu H.-B.,  2009, \mn@doi [JCAP]
  {10.1088/1475-7516/2009/07/004}, \href
  {https://ui.adsabs.harvard.edu/abs/2009JCAP...07..004F} {2009, 004}

\bibitem[\protect\citeauthoryear{Fischer, Br{\"u}ggen, {Schmidt-Hoberg}, Dolag,
  Kahlhoefer, Ragagnin  \& Robertson}{Fischer
  et~al.}{2021}]{fischerNbodySimulationsDark2021a}
Fischer M.~S.,  Br{\"u}ggen M.,  {Schmidt-Hoberg} K.,  Dolag K.,  Kahlhoefer
  F.,  Ragagnin A.,   Robertson A.,  2021, \mn@doi [MNRAS]
  {10.1093/mnras/stab1198}, \href
  {https://ui.adsabs.harvard.edu/abs/2021MNRAS.505..851F} {505, 851}

\bibitem[\protect\citeauthoryear{Fischer, Br{\"u}ggen, {Schmidt-Hoberg}, Dolag,
  Ragagnin  \& Robertson}{Fischer
  et~al.}{2022}]{fischerUnequalmassMergersDark2022}
Fischer M.~S.,  Br{\"u}ggen M.,  {Schmidt-Hoberg} K.,  Dolag K.,  Ragagnin A.,
   Robertson A.,  2022, \mn@doi [MNRAS] {10.1093/mnras/stab3544}, \href
  {https://ui.adsabs.harvard.edu/abs/2022MNRAS.510.4080F} {510, 4080}

\bibitem[\protect\citeauthoryear{Fischer, Kasselmann, Brüggen, Dolag,
  Kahlhoefer, Ragagnin, Robertson  \& Schmidt-Hoberg}{Fischer
  et~al.}{2023a}]{fischer2023d}
Fischer M.~S.,  Kasselmann L.,  Brüggen M.,  Dolag K.,  Kahlhoefer F.,
  Ragagnin A.,  Robertson A.,   Schmidt-Hoberg K.,  2023a, Cosmological and
  idealised simulations of dark matter haloes with velocity-dependent, rare and
  frequent self-interactions (\mn@eprint {arXiv} {2310.07750})

\bibitem[\protect\citeauthoryear{Fischer, Durke, Hollingshausen, Hammer,
  Br{\"u}ggen  \& Dolag}{Fischer
  et~al.}{2023b}]{fischerRoleBaryonsSelfinteracting2023}
Fischer M.~S.,  Durke N.-H.,  Hollingshausen K.,  Hammer C.,  Br{\"u}ggen M.,
  Dolag K.,  2023b, \mn@doi [MNRAS] {10.1093/mnras/stad1786}, \href
  {https://ui.adsabs.harvard.edu/abs/2023MNRAS.523.5915F} {523, 5915}

\bibitem[\protect\citeauthoryear{Gilman, Bovy, Treu, Nierenberg, Birrer, Benson
   \& Sameie}{Gilman et~al.}{2021}]{gilmanStrongLensingSignatures2021}
Gilman D.,  Bovy J.,  Treu T.,  Nierenberg A.,  Birrer S.,  Benson A.,   Sameie
  O.,  2021, \mn@doi [MNRAS] {10.1093/mnras/stab2335}, \href
  {https://ui.adsabs.harvard.edu/abs/2021MNRAS.507.2432G} {507, 2432}

\bibitem[\protect\citeauthoryear{Harvey, Massey, Kitching, Taylor  \&
  Tittley}{Harvey et~al.}{2015}]{harveyNongravitationalInteractionsDark2015}
Harvey D.,  Massey R.,  Kitching T.,  Taylor A.,   Tittley E.,  2015, \mn@doi
  [Sci] {10.1126/science.1261381}, \href
  {https://ui.adsabs.harvard.edu/abs/2015Sci...347.1462H} {347, 1462}

\bibitem[\protect\citeauthoryear{Harvey, Robertson, Massey  \& McCarthy}{Harvey
  et~al.}{2019}]{harveyObservableTestsSelfinteracting2019}
Harvey D.,  Robertson A.,  Massey R.,   McCarthy I.~G.,  2019, \mn@doi [MNRAS]
  {10.1093/mnras/stz1816}, \href
  {https://ui.adsabs.harvard.edu/abs/2019MNRAS.488.1572H} {488, 1572}

\bibitem[\protect\citeauthoryear{Jee, Hughes, Menanteau, Sif{\'o}n, Mandelbaum,
  Barrientos, Infante  \& Ng}{Jee et~al.}{2014}]{jeeWeighingGordoPrecision2014}
Jee M.~J.,  Hughes J.~P.,  Menanteau F.,  Sif{\'o}n C.,  Mandelbaum R.,
  Barrientos L.~F.,  Infante L.,   Ng K.~Y.,  2014, \mn@doi [ApJ]
  {10.1088/0004-637X/785/1/20}, \href
  {https://ui.adsabs.harvard.edu/abs/2014ApJ...785...20J} {785, 20}

\bibitem[\protect\citeauthoryear{Jee et~al.,}{Jee
  et~al.}{2015}]{jeeMCConstrainingDark2015}
Jee M.~J.,  et~al., 2015, \mn@doi [ApJ] {10.1088/0004-637X/802/1/46}, \href
  {https://ui.adsabs.harvard.edu/abs/2015ApJ...802...46J} {802, 46}

\bibitem[\protect\citeauthoryear{Kahlhoefer, {Schmidt-Hoberg}, Frandsen  \&
  Sarkar}{Kahlhoefer et~al.}{2014}]{kahlhoeferCollidingClustersDark2014}
Kahlhoefer F.,  {Schmidt-Hoberg} K.,  Frandsen M.~T.,   Sarkar S.,  2014,
  \mn@doi [MNRAS] {10.1093/mnras/stt2097}, \href
  {https://ui.adsabs.harvard.edu/abs/2014MNRAS.437.2865K} {437, 2865}

\bibitem[\protect\citeauthoryear{Kahlhoefer, {Schmidt-Hoberg}  \&
  Wild}{Kahlhoefer et~al.}{2017}]{kahlhoeferDarkMatterSelfinteractions2017}
Kahlhoefer F.,  {Schmidt-Hoberg} K.,   Wild S.,  2017, \mn@doi [JCAP]
  {10.1088/1475-7516/2017/08/003}, \href {http://arxiv.org/abs/1704.02149}
  {2017, 003}

\bibitem[\protect\citeauthoryear{Kim, Peter  \& Wittman}{Kim
  et~al.}{2017}]{kimWakeDarkGiants2017}
Kim S.~Y.,  Peter A. H.~G.,   Wittman D.,  2017, \mn@doi [MNRAS]
  {10.1093/mnras/stx896}, \href
  {https://ui.adsabs.harvard.edu/abs/2017MNRAS.469.1414K} {469, 1414}

\bibitem[\protect\citeauthoryear{Kummer, Kahlhoefer  \&
  {Schmidt-Hoberg}}{Kummer et~al.}{2018}]{kummerEffectiveDescriptionDark2018}
Kummer J.,  Kahlhoefer F.,   {Schmidt-Hoberg} K.,  2018, \mn@doi [MNRAS]
  {10.1093/mnras/stx2715}, \href
  {https://ui.adsabs.harvard.edu/abs/2018MNRAS.474..388K} {474, 388}

\bibitem[\protect\citeauthoryear{Kummer, Br{\"u}ggen, Dolag, Kahlhoefer  \&
  {Schmidt-Hoberg}}{Kummer et~al.}{2019}]{kummerSimulationsCoreFormation2019}
Kummer J.,  Br{\"u}ggen M.,  Dolag K.,  Kahlhoefer F.,   {Schmidt-Hoberg} K.,
  2019, \mn@doi [MNRAS] {10.1093/mnras/stz1261}, \href
  {https://ui.adsabs.harvard.edu/abs/2019MNRAS.487..354K} {487, 354}

\bibitem[\protect\citeauthoryear{Lauer, Postman, Strauss, Graves  \&
  Chisari}{Lauer et~al.}{2014}]{lauerBrightestClusterGalaxies2014}
Lauer T.~R.,  Postman M.,  Strauss M.~A.,  Graves G.~J.,   Chisari N.~E.,
  2014, \mn@doi [ApJ] {10.1088/0004-637X/797/2/82}, \href
  {https://ui.adsabs.harvard.edu/abs/2014ApJ...797...82L} {797, 82}

\bibitem[\protect\citeauthoryear{Laureijs et~al.,}{Laureijs
  et~al.}{2011}]{laureijsEuclidDefinitionStudy2011a}
Laureijs R.,  et~al., 2011, Euclid {{Definition Study Report}} (\mn@eprint {}
  {1110.3193})

\bibitem[\protect\citeauthoryear{Navarro, Frenk  \& White}{Navarro
  et~al.}{1996}]{navarroStructureColdDark1996a}
Navarro J.~F.,  Frenk C.~S.,   White S. D.~M.,  1996, \mn@doi [ApJ]
  {10.1086/177173}, \href
  {https://ui.adsabs.harvard.edu/abs/1996ApJ...462..563N} {462, 563}

\bibitem[\protect\citeauthoryear{O'Neil et~al.,}{O'Neil
  et~al.}{2023}]{oneilEndothermicSelfinteractingDark2023}
O'Neil S.,  et~al., 2023, \mn@doi [MNRAS] {10.1093/mnras/stad1850}, \href
  {https://ui.adsabs.harvard.edu/abs/2023MNRAS.524..288O} {524, 288}

\bibitem[\protect\citeauthoryear{Outmezguine, Boddy, {Gad-Nasr}, Kaplinghat  \&
  Sagunski}{Outmezguine
  et~al.}{2023}]{outmezguineUniversalGravothermalEvolution2023}
Outmezguine N.~J.,  Boddy K.~K.,  {Gad-Nasr} S.,  Kaplinghat M.,   Sagunski L.,
   2023, \mn@doi [MNRAS] {10.1093/mnras/stad1705}, \href
  {https://ui.adsabs.harvard.edu/abs/2023MNRAS.523.4786O} {523, 4786}

\bibitem[\protect\citeauthoryear{Randall, Markevitch, Clowe, Gonzalez  \&
  Brada{\v c}}{Randall
  et~al.}{2008}]{randallConstraintsSelfInteractionCross2008}
Randall S.~W.,  Markevitch M.,  Clowe D.,  Gonzalez A.~H.,   Brada{\v c} M.,
  2008, \mn@doi [ApJ] {10.1086/587859}, \href
  {https://ui.adsabs.harvard.edu/abs/2008ApJ...679.1173R} {679, 1173}

\bibitem[\protect\citeauthoryear{Ren, Kwa, Kaplinghat  \& Yu}{Ren
  et~al.}{2019}]{renReconcilingDiversityUniformity2019}
Ren T.,  Kwa A.,  Kaplinghat M.,   Yu H.-B.,  2019, \mn@doi [PhRvX]
  {10.1103/PhysRevX.9.031020}, \href
  {https://ui.adsabs.harvard.edu/abs/2019PhRvX...9c1020R} {9, 031020}

\bibitem[\protect\citeauthoryear{Robertson, Massey  \& Eke}{Robertson
  et~al.}{2017a}]{robertsonWhatDoesBullet2017}
Robertson A.,  Massey R.,   Eke V.,  2017a, \mn@doi [MNRAS]
  {10.1093/mnras/stw2670}, \href
  {https://ui.adsabs.harvard.edu/abs/2017MNRAS.465..569R} {465, 569}

\bibitem[\protect\citeauthoryear{Robertson, Massey  \& Eke}{Robertson
  et~al.}{2017b}]{robertsonCosmicParticleColliders2017}
Robertson A.,  Massey R.,   Eke V.,  2017b, \mn@doi [MNRAS]
  {10.1093/mnras/stx463}, \href
  {https://ui.adsabs.harvard.edu/abs/2017MNRAS.467.4719R} {467, 4719}

\bibitem[\protect\citeauthoryear{Rocha, Peter, Bullock, Kaplinghat,
  {Garrison-Kimmel}, O{\~n}orbe  \& Moustakas}{Rocha
  et~al.}{2013}]{rochaCosmologicalSimulationsSelfinteracting2013}
Rocha M.,  Peter A. H.~G.,  Bullock J.~S.,  Kaplinghat M.,  {Garrison-Kimmel}
  S.,  O{\~n}orbe J.,   Moustakas L.~A.,  2013, \mn@doi [MNRAS]
  {10.1093/mnras/sts514}, \href
  {https://ui.adsabs.harvard.edu/abs/2013MNRAS.430...81R} {430, 81}

\bibitem[\protect\citeauthoryear{Romualdez et~al.,}{Romualdez
  et~al.}{2016}]{romualdezDesignDevelopmentHighresolution2016}
Romualdez L.~J.,  et~al., 2016, The Design and Development of a High-Resolution
  Visible-to-near-{{UV}} Telescope for Balloon-Borne Astronomy: {{SuperBIT}}
  (\mn@eprint {} {1608.02502})

\bibitem[\protect\citeauthoryear{Sagunski, {Gad-Nasr}, Colquhoun, Robertson  \&
  Tulin}{Sagunski
  et~al.}{2021}]{sagunskiVelocitydependentSelfinteractingDark2021}
Sagunski L.,  {Gad-Nasr} S.,  Colquhoun B.,  Robertson A.,   Tulin S.,  2021,
  \mn@doi [JCAP] {10.1088/1475-7516/2021/01/024}, \href
  {https://ui.adsabs.harvard.edu/abs/2021JCAP...01..024S} {2021, 024}

\bibitem[\protect\citeauthoryear{Sankar~Ray, Sarkar  \& Kumar~Shaw}{Sankar~Ray
  et~al.}{2022}]{sankarrayConstraintsDarkMatter2022}
Sankar~Ray T.,  Sarkar S.,   Kumar~Shaw A.,  2022, \mn@doi [JCAP]
  {10.1088/1475-7516/2022/09/011}, \href
  {https://ui.adsabs.harvard.edu/abs/2022JCAP...09..011S} {2022, 011}

\bibitem[\protect\citeauthoryear{Spergel \& Steinhardt}{Spergel \&
  Steinhardt}{2000}]{spergelObservationalEvidenceSelfInteracting2000}
Spergel D.~N.,  Steinhardt P.~J.,  2000, \mn@doi [PhRvL]
  {10.1103/PhysRevLett.84.3760}, \href
  {https://ui.adsabs.harvard.edu/abs/2000PhRvL..84.3760S} {84, 3760}

\bibitem[\protect\citeauthoryear{Tulin \& Yu}{Tulin \&
  Yu}{2018}]{tulinDarkMatterSelfinteractions2018}
Tulin S.,  Yu H.-B.,  2018, \mn@doi [PhR] {10.1016/j.physrep.2017.11.004},
  \href {https://ui.adsabs.harvard.edu/abs/2018PhR...730....1T} {730, 1}

\bibitem[\protect\citeauthoryear{Tulin, Yu  \& Zurek}{Tulin
  et~al.}{2013}]{tulincollisionlessDarkMatter2013}
Tulin S.,  Yu H.-B.,   Zurek K.~M.,  2013, \mn@doi [PhRvD]
  {10.1103/PhysRevD.87.115007}, \href
  {https://ui.adsabs.harvard.edu/abs/2013PhRvD..87k5007T} {87, 115007}

\bibitem[\protect\citeauthoryear{Yang \& Yu}{Yang \&
  Yu}{2022}]{yangGravothermalEvolutionDark2022}
Yang D.,  Yu H.-B.,  2022, \mn@doi [JCAP] {10.1088/1475-7516/2022/09/077},
  \href {https://ui.adsabs.harvard.edu/abs/2022JCAP...09..077Y} {2022, 077}

\bibitem[\protect\citeauthoryear{Yang, Du, Zeng, Benson, Jiang, Nadler  \&
  Peter}{Yang et~al.}{2023}]{yangGravothermalSolutionsSIDM2023}
Yang S.,  Du X.,  Zeng Z.~C.,  Benson A.,  Jiang F.,  Nadler E.~O.,   Peter A.
  H.~G.,  2023, \mn@doi [ApJ] {10.3847/1538-4357/acbd49}, \href
  {https://ui.adsabs.harvard.edu/abs/2023ApJ...946...47Y} {946, 47}

\makeatother
\end{thebibliography}




\appendix
\section{Validating lower resolution simulations}\label{app:validating resolution}
\begin{figure}
    \centering
    \includegraphics[width=0.48\textwidth]{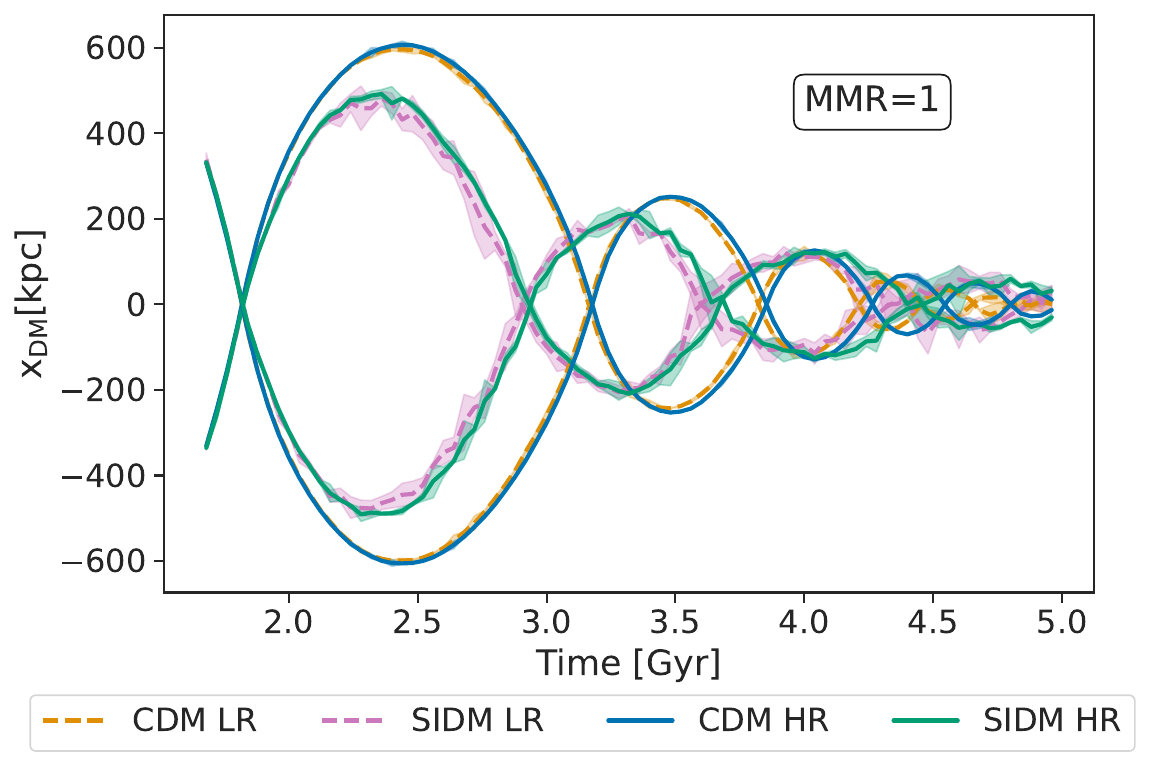}
    \includegraphics[width=0.48\textwidth]{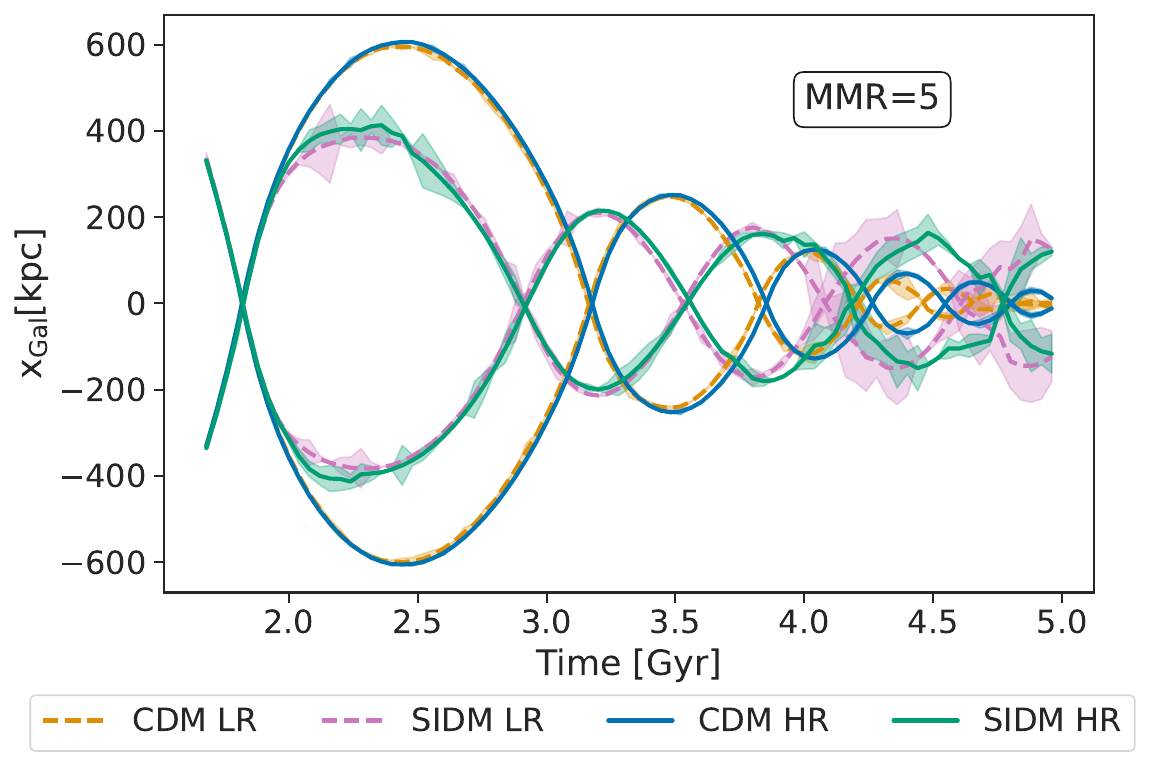}
    \includegraphics[width=0.48\textwidth]{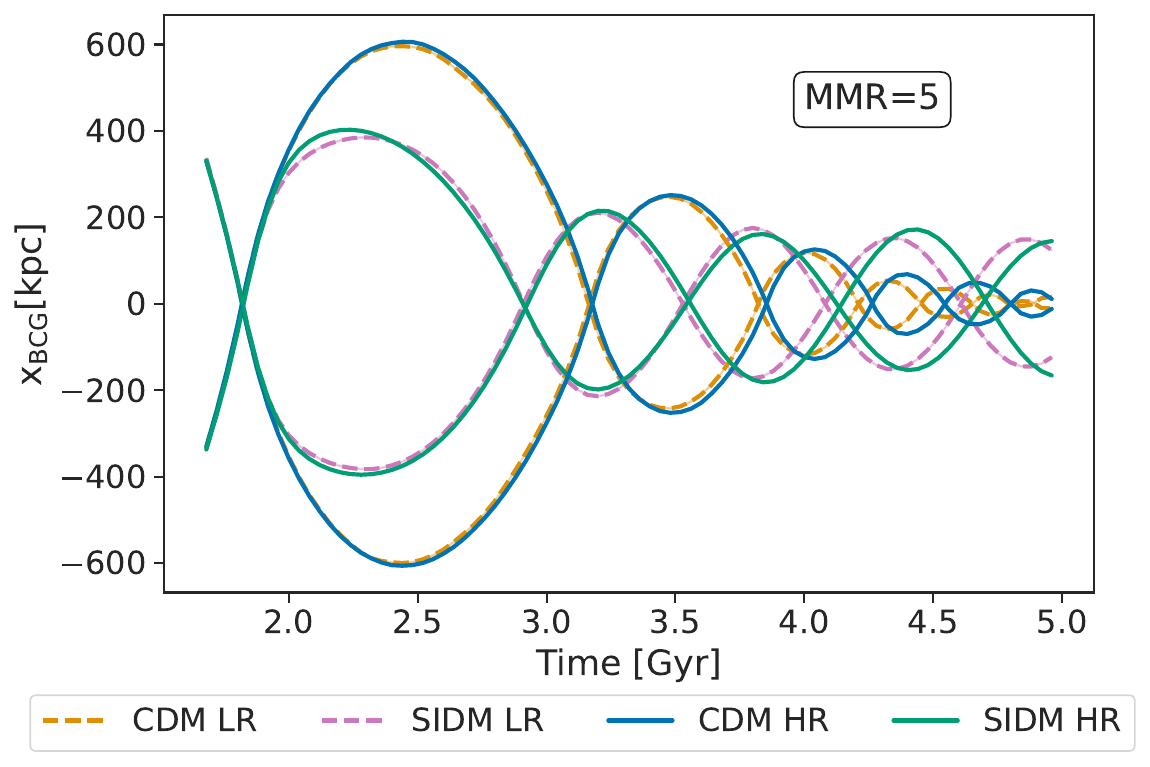}
    \caption{Comparison of peaks positions of different components between low and high resolutions for both CDM and SIDM simulations in the equal mass merger. Top, middle and bottom panels correspond DM, galaxy and BCG components. The dashed lines correspond to low resolution and solid lines correspond to high resolution. The SIDM case corresponds to the frequent self-interactions with $\som = \SI{0.5}{\m\squared\per\g}$.}
    \label{fig:hr vs lr m5r1}
\end{figure}

\begin{figure}
    \centering
    \includegraphics[width=0.48\textwidth]{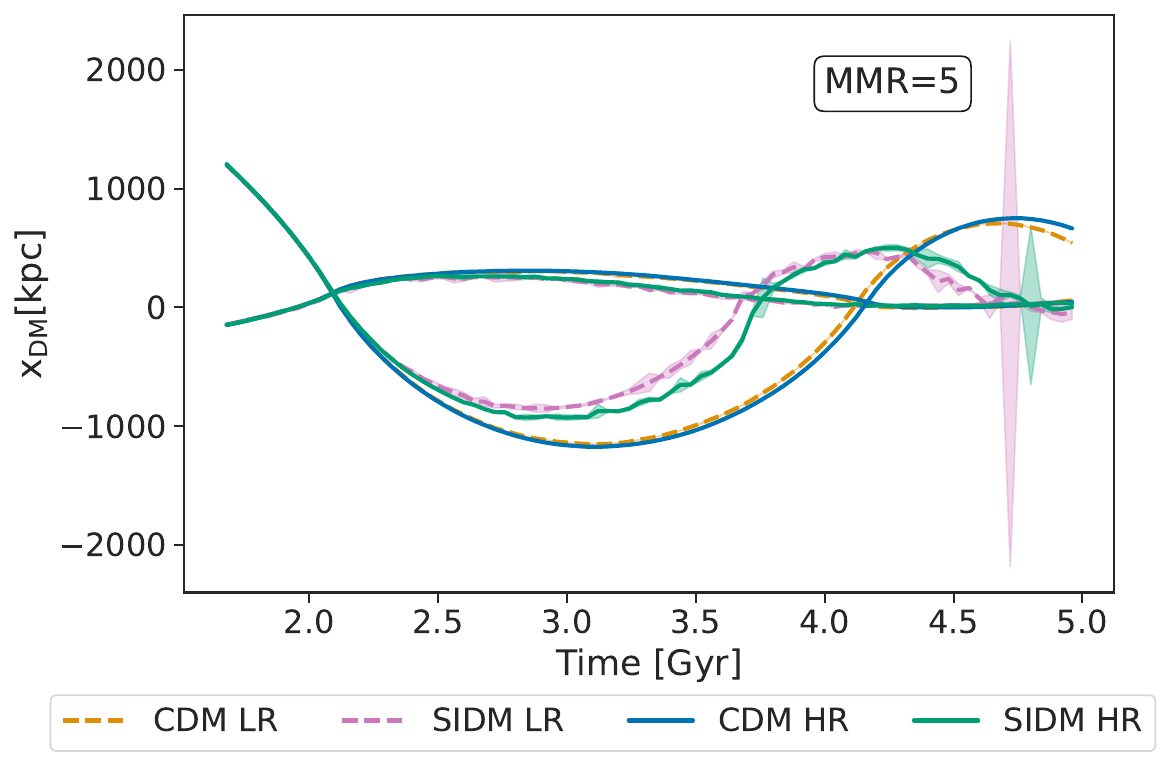}
    \includegraphics[width=0.48\textwidth]{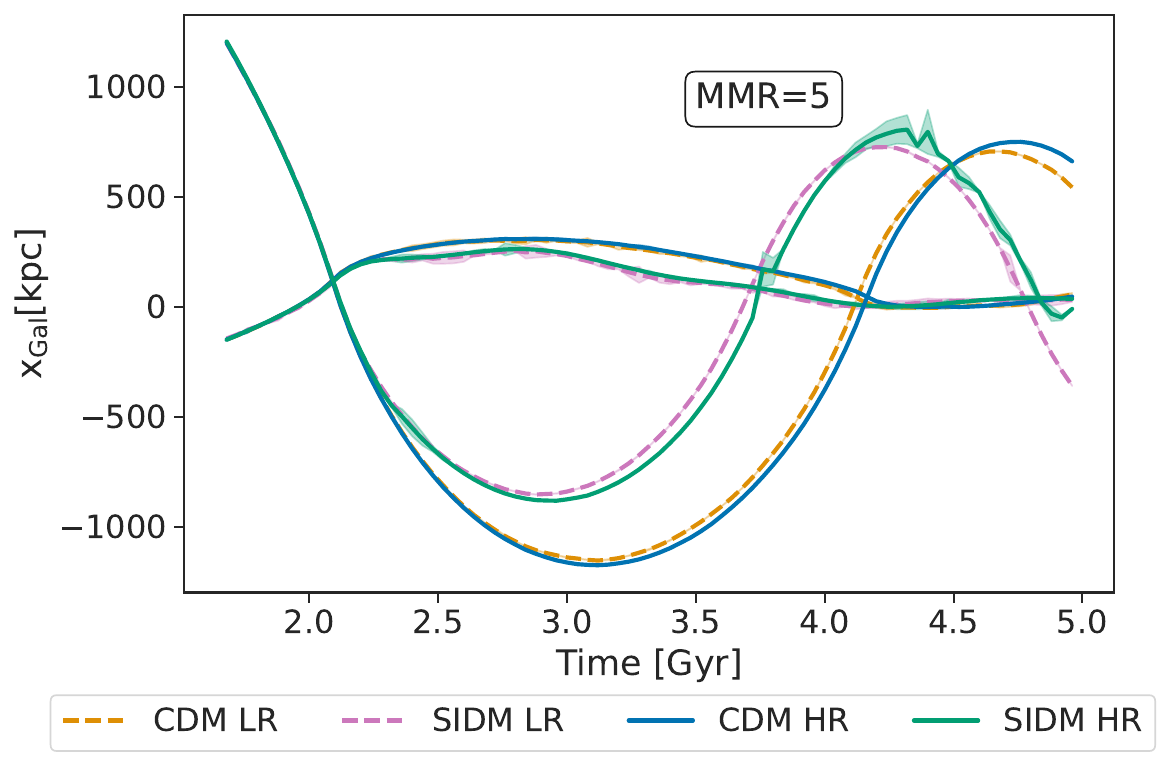}
    \includegraphics[width=0.48\textwidth]{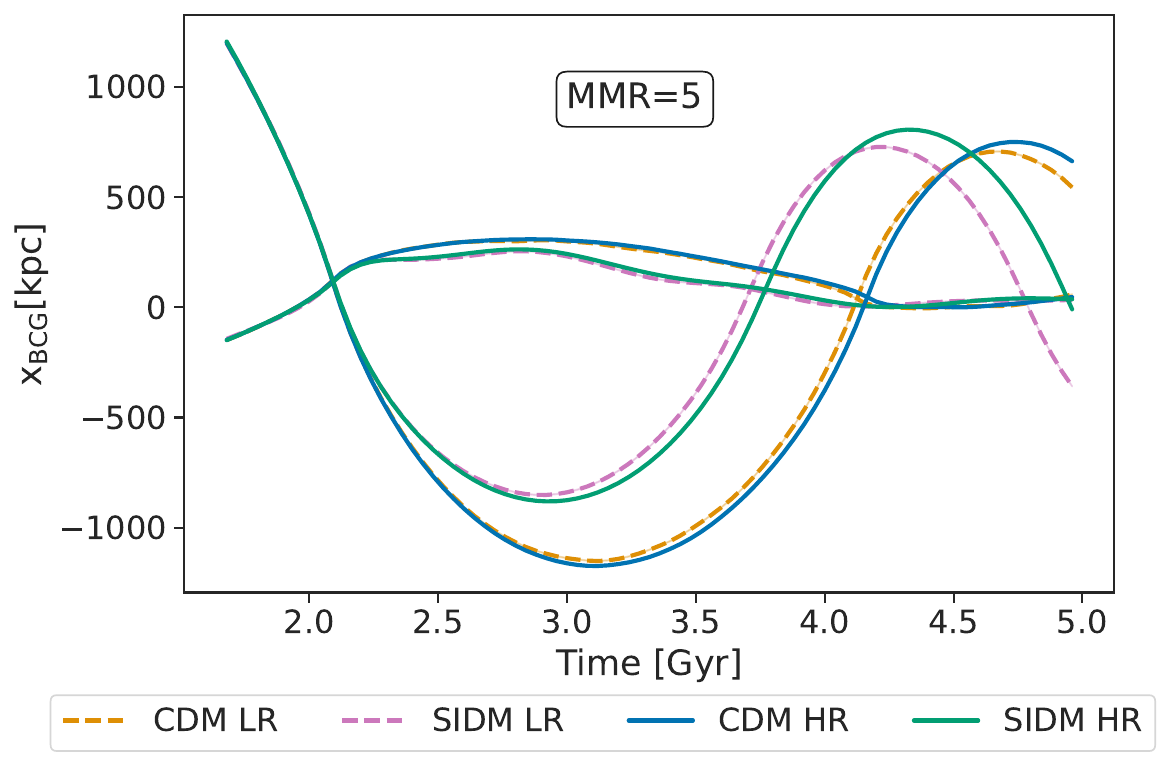}
    \caption{Comparison of peaks positions of different components between low and high resolutions for CDM and SIDM simulations in the unequal mass merger. Top, middle and bottom panels correspond DM, galaxy and BCG components. The dashed lines correspond to low resolution and solid lines correspond to high resolution. The SIDM case corresponds to the frequent self-interactions with $\som = \SI{0.5}{\m\squared\per\g}$.}
    \label{fig:hr vs lr m5r5}
\end{figure}

In this section, we compare the peak positions of DM, galaxy and BCG components between low resolution and high resolution simulations. The low- and the high-resolution simulation use the NFW parameters given in \autoref{tab:merger nfw params} for generating the haloes. The only difference being that the DM and galaxy particles in the high-resolution simulation have a resolution of $10^7$ particles instead of $10^6$ particles. Both of them also use the same initial conditions as given in \autoref{tab:merger ics}. The DM component is simulated with and without self-interactions. For the SIDM case,  we simulate with frequent self-interactions with a \review{constant $\sigma_T$ of \SI{0.5}{\cm\squared\per\g}.} We observe that the peak positions evolve almost identically independent of the resolution up until 5 billion years. See 
\autoref{fig:hr vs lr m5r1} and \ref{fig:hr vs lr m5r5}.

\section{Testing rescaling}\label{app:central density rescaling}
\begin{figure}
    \centering
    \includegraphics[width=0.45\textwidth]{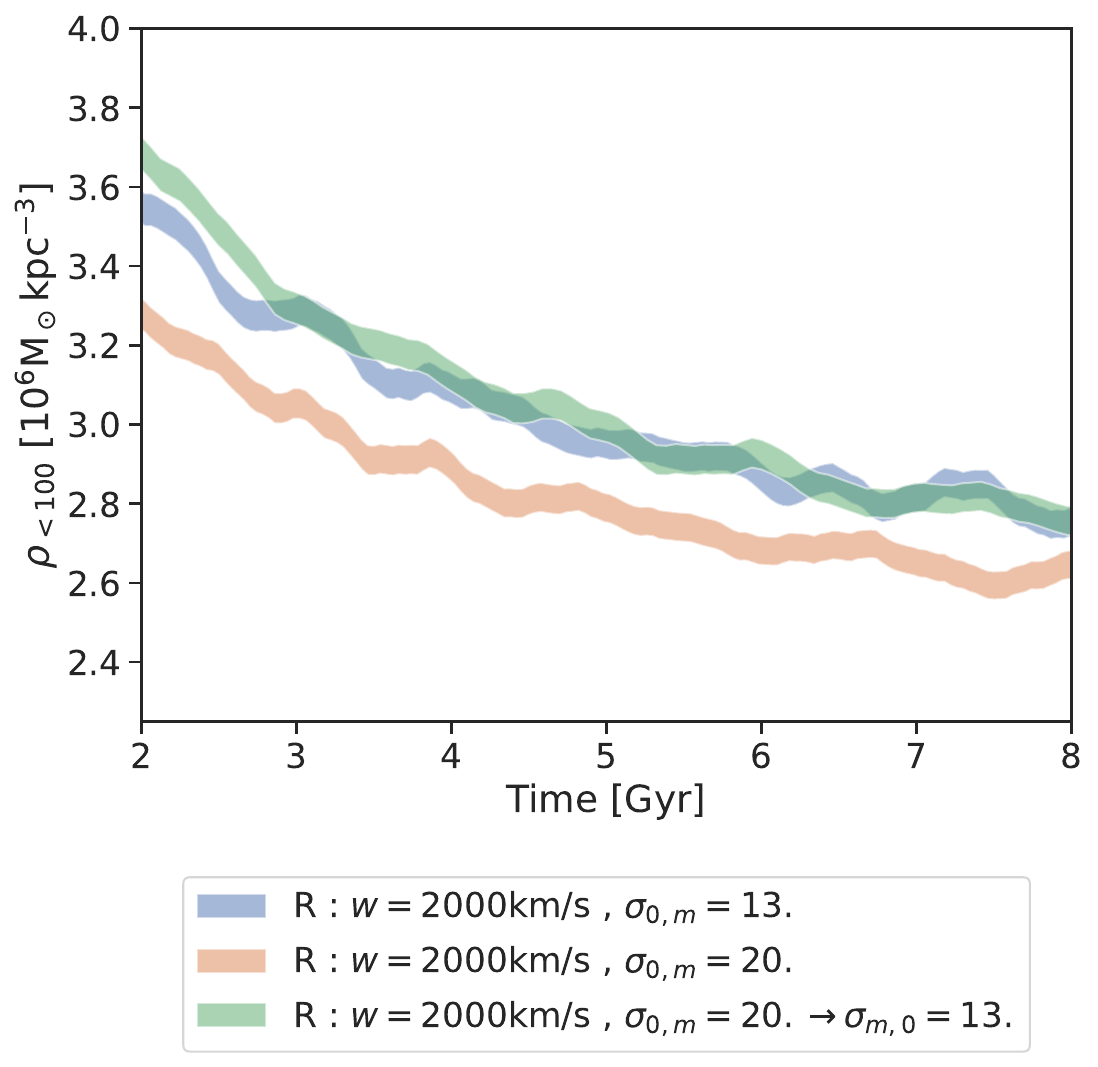}
    \includegraphics[width=0.45\textwidth]{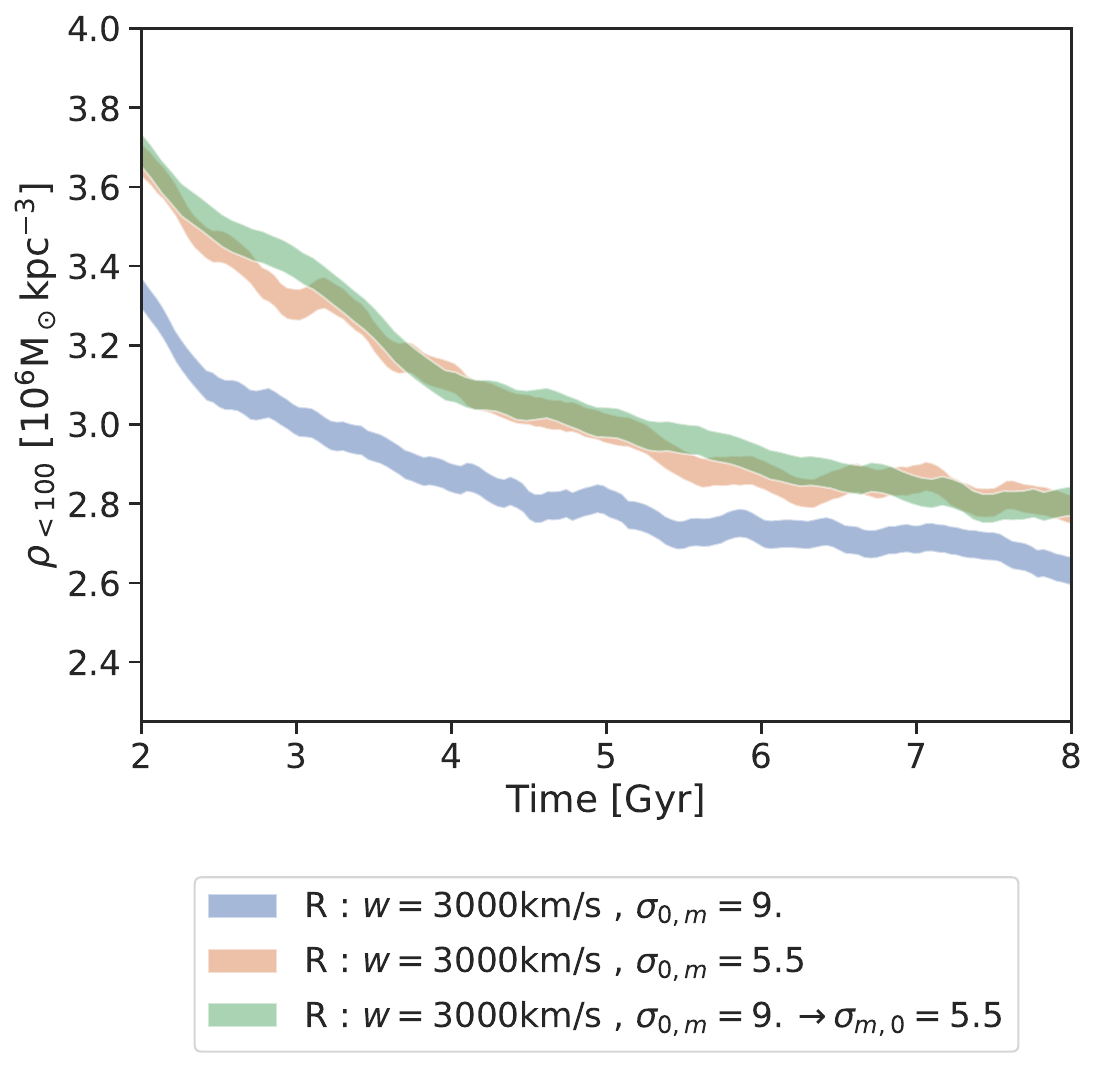}
    \caption{Evolution of central density for two values of $w$,  \SI{2000}{\km\per\s} and \SI{3000}{\km\per\s} in the top and bottom panel, respectively.}
    \label{fig:appendix rescaling}
\end{figure}

We test the rescaling by $\som$ for a given $w$ with rare self-interactions. \autoref{fig:appendix rescaling} shows the evolution of central density of an isolated halo for two values of $w$ -- \SI{2000}{\km\per\s} and \SI{3000}{\km\per\s} -- in the left and right panel, respectively. For example, in the left panel, after rescaling $t$ of $\som = $ \SI{20}{\cm\squared\per\g} simulation by a factor 20/13, the evolution is similar to the simulation with $\som=\SI{13}{\cm\squared\per\g}$.

\section{Central density evolution in Merger}\label{app:cd evolution in merger}

\begin{figure}
    \centering
    \includegraphics[width=0.48\textwidth]{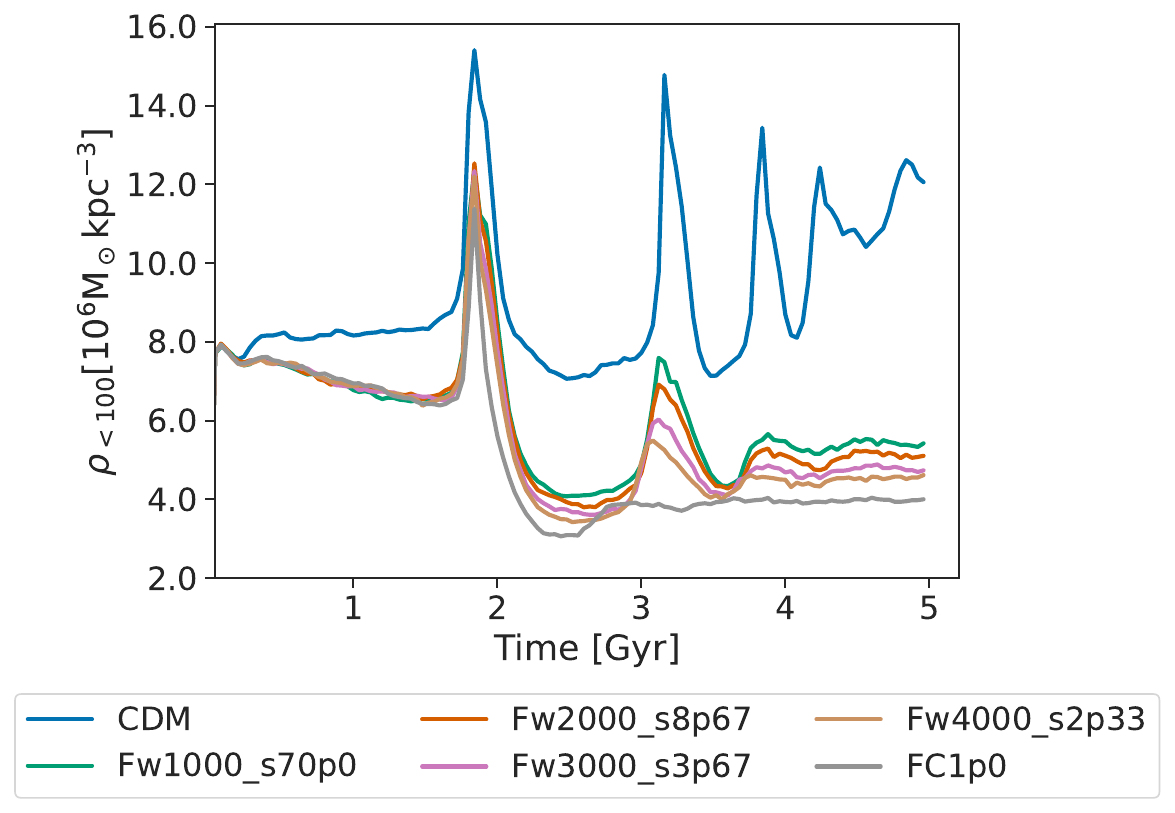}
    \caption{Evolution of the central density measured within \SI{100}{\kpc} around the DM peak of main halo. The parameters are CD-matched and the labels are explained in \autoref{tab:seff 1.5 labels}.}
    \label{fig:M5R1 central density seff 1.5}
\end{figure}

\begin{figure}
    \centering
    \includegraphics[width=0.48\textwidth]{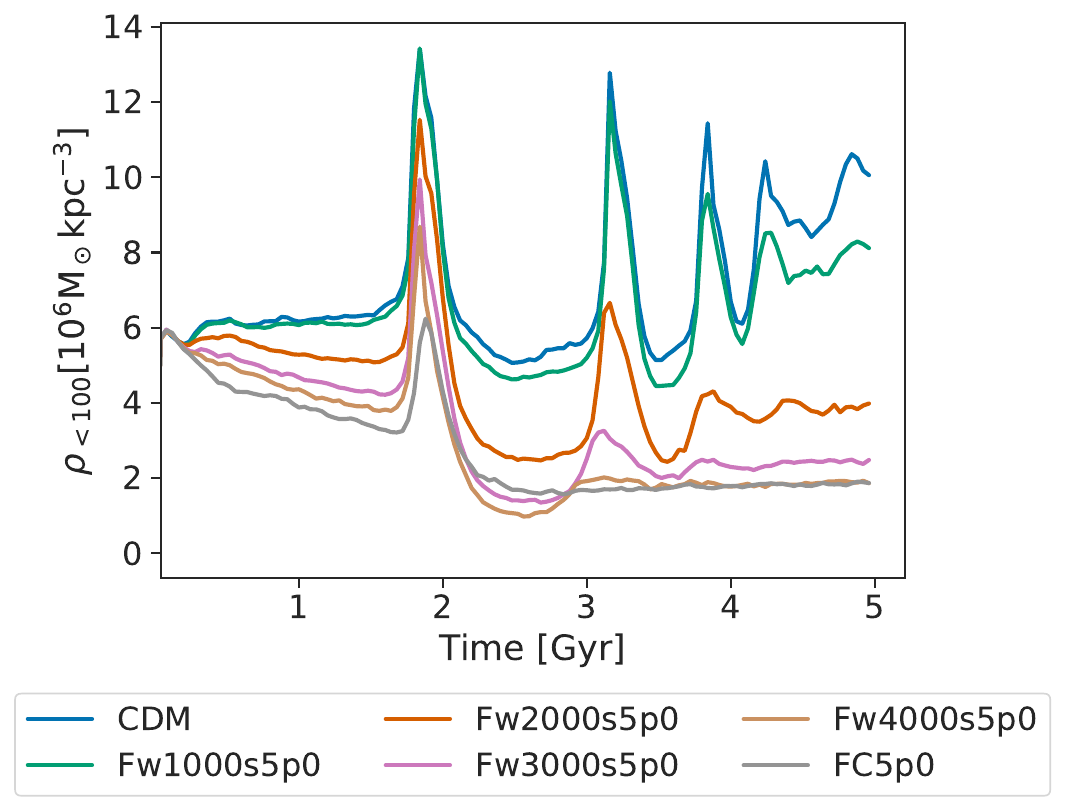}
    \caption{Evolution of the central density measured within \SI{100}{\kpc}, around the DM peak of main halo \review{in the equal mass merger}. The parameters have the same value of $\som=\SI{5}{\cm\squared\per\g}$, but with varying values of $w$. The labels are explained in \autoref{tab:s5p0 labels}.}
    \label{fig:M5R1 central density s5p0}
\end{figure}

In \autoref{fig:M5R1 central density seff 1.5}, we show the evolution of the central density around the DM peak of the main halo \review{in the equal mass merger}. The curves correspond to simulations where the cross-section parameters are CD-matched. Similarly, in \autoref{fig:M5R1 central density s5p0} we show the central densities for cases when the cross-section parameters have the same $\som$, but with varying $w$. 

\begin{figure}
        \centering
        \includegraphics[width=0.48\textwidth]{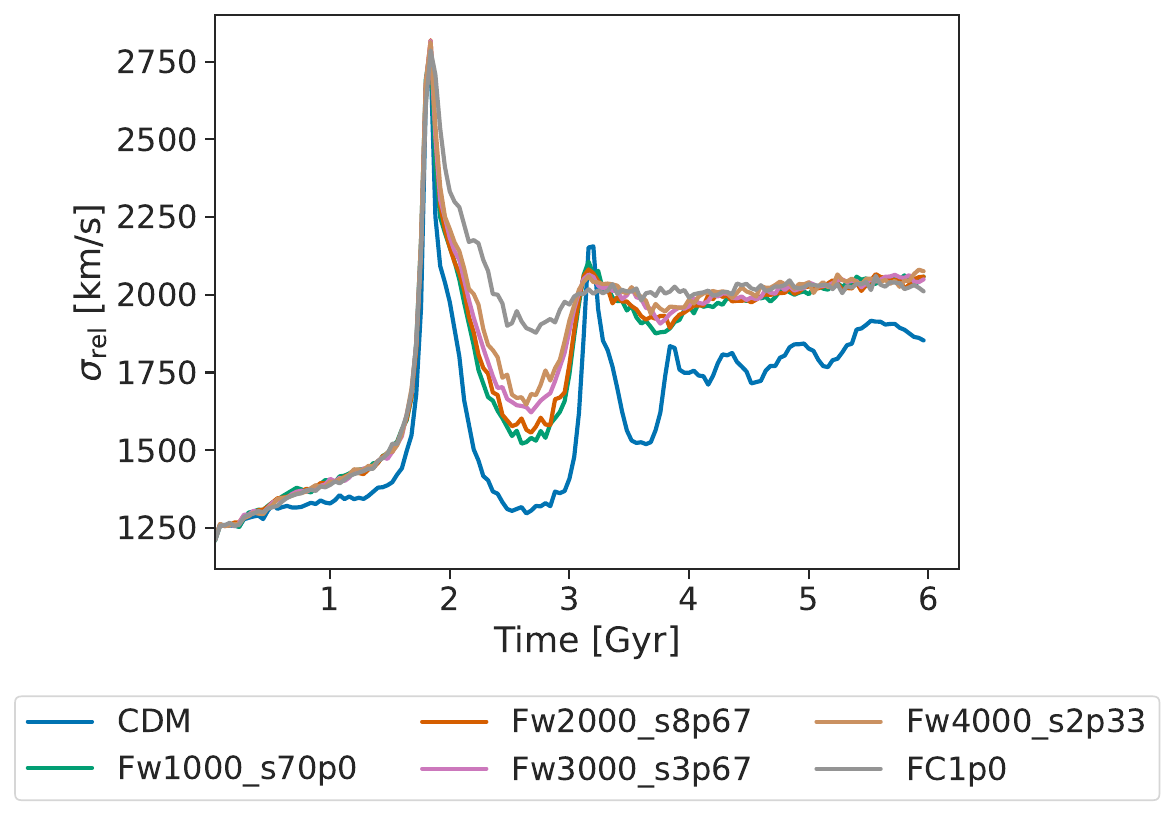}
        \caption{\review{Evolution of the central relative velocity dispersion measured within \SI{100}{\kpc} around the DM peak of main halo. The parameters are CD-matched and the labels are explained in \autoref{tab:seff 1.5 labels}.}
        \label{fig:M5R1 veldispr seff1p0}}
    \end{figure}

\begin{figure}
        \centering
        \includegraphics[width=0.48\textwidth]{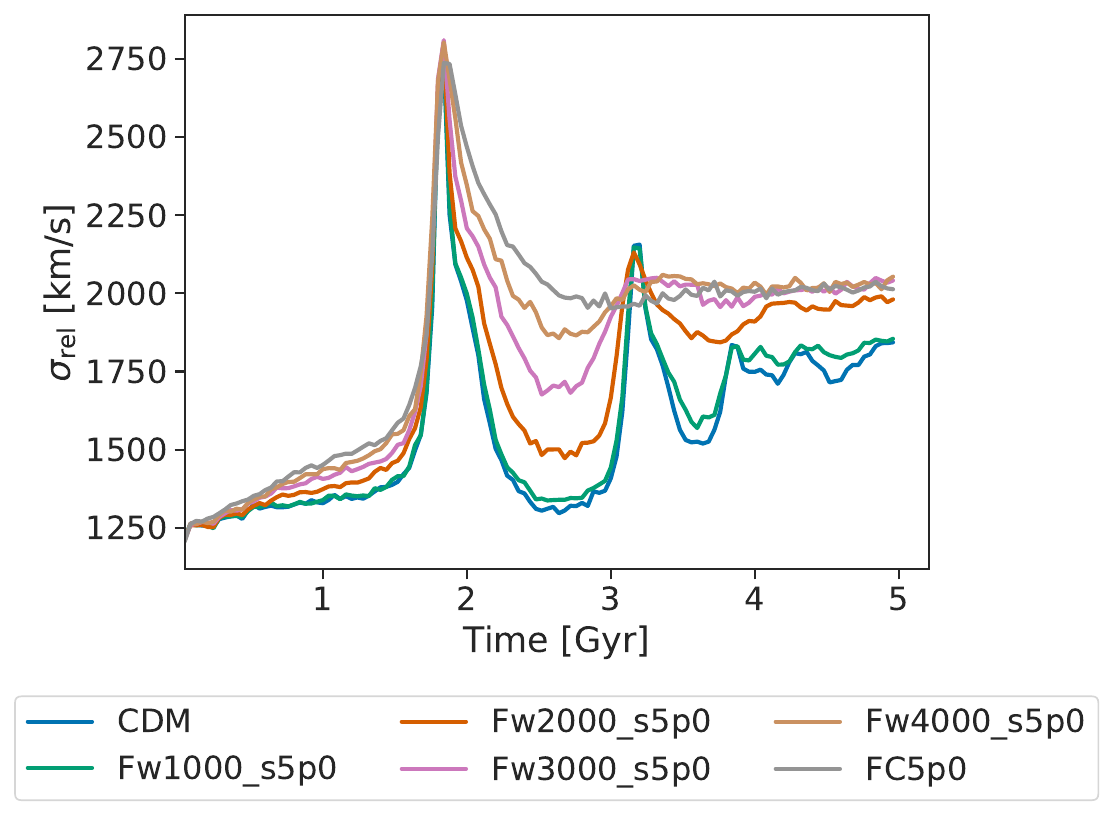}
        \caption{\review{Evolution of the central relative velocity dispersion within \SI{100}{\kpc} around the DM peak of the main halo. The parameters have the same value of $\som=\SI{5}{\cm\squared\per\g}$, but with varying values of $w$. The labels are explained in \autoref{tab:s5p0 labels}.}
        \label{fig:M5R1 veldispr s5p0}}
    \end{figure}

\review{
\section{Central velocity dispersion evolution in Merger}\label{ap:veldispr evolution in merger}

The relative velocity dispersion around the DM peak within \SI{100}{\kpc} \review{in the equal mass merger} is shown as a function of time in \autoref{fig:M5R1 veldispr seff1p0} and \autoref{fig:M5R1 veldispr s5p0}. The earlier figure corresponds to simulations with cross-section parameters that are CD-matched, while the latter figure corresponds to cross-section parameters with fixed $\som$ and varying $w$. The relative velocity dispersion is calculated from the 1D velocity dispersion, i.e. $\sigma_{\rm rel} = \sqrt{2} \sigma_{1D}$. }

\review{
\section{Momentum transfer cross-section of CD-matched parameters}
Plots similar to \autoref{fig:cross-section} but with CD-matched parameters are provided. \autoref{fig:cross-section-seff} and \ref{fig:cross-section-upperBound} correspond to the parameters in \autoref{tab:seff 1.5 labels} and \ref{tab:upper bound params}, respectively. 
\begin{figure}
    \centering
    \includegraphics[width=0.48\textwidth]{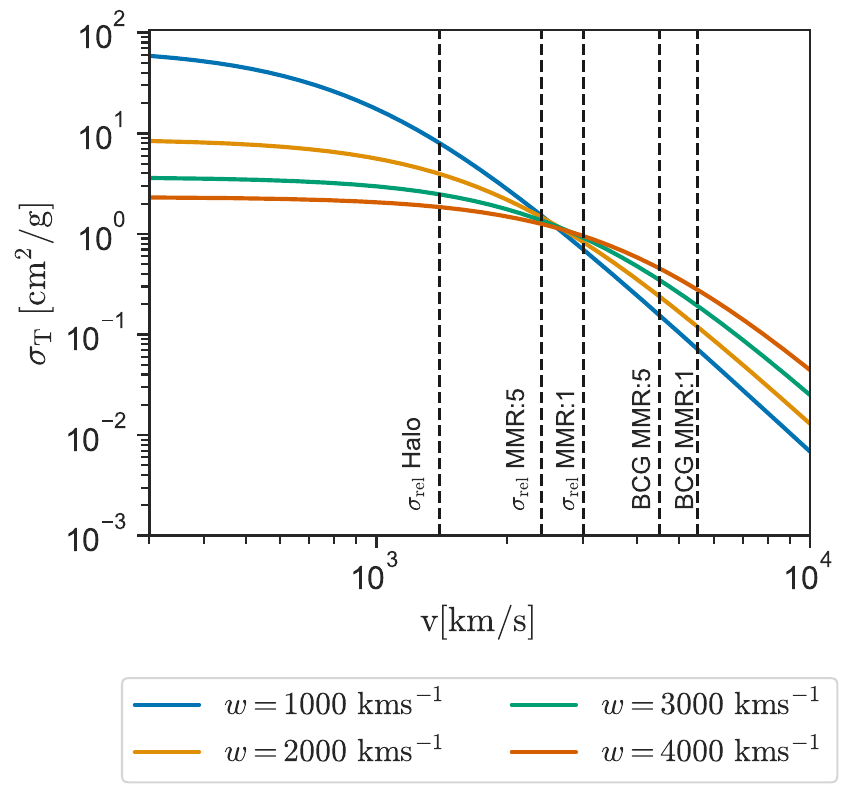}
    \caption{Momentum transfer cross-section as a function of velocity for the parameters given in \autoref{tab:seff 1.5 labels}. The vertical dashed lines are explained in the captions of \autoref{fig:cross-section}.}
    \label{fig:cross-section-seff}
\end{figure}

\begin{figure}
    \centering
    \includegraphics[width=0.48\textwidth]{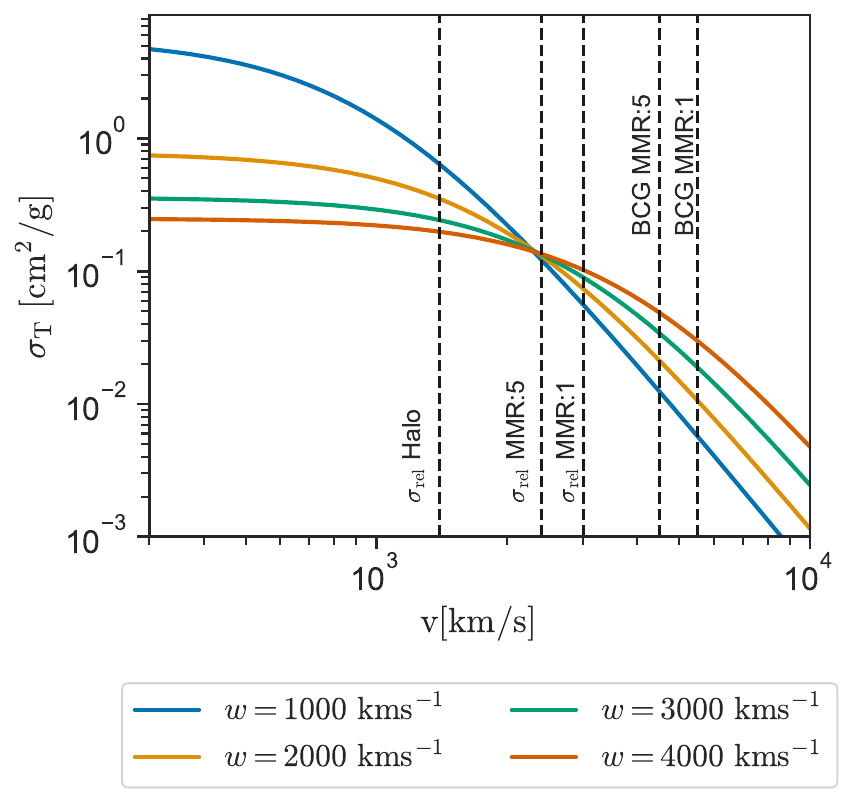}
    \caption{Momentum transfer cross-section as a function of velocity for the parameters given in \autoref{tab:upper bound params}.The vertical dashed lines are explained in the captions of \autoref{fig:cross-section}.}
    \label{fig:cross-section-upperBound}
\end{figure}
}
\bsp	
\label{lastpage}
\end{document}